\documentclass[]{aa}
 \usepackage{graphicx}
 \def\url#1{{\bf #1}}
 \def\figw{9cm}

 \def\m@th{\mathsurround=0pt}
 \def\EQM#1{\vcenter{\normalbaselines\m@th
     \ialign{${\displaystyle ##}$\hfil&&\ ${\displaystyle ##}$\hfil\crcr
     \mathstrut\crcr\noalign{\kern-\baselineskip}
     \noalign{\smallskip}
     #1\crcr\mathstrut\crcr\noalign{\kern-\baselineskip}}}}

 \def\aj{AJ}%
 \def\apjl{ApJ}%
 \def\apjs{ApJS}%
 \def\ao{Appl.~Opt.}%
 \def\aap{A\&A}%
 \def\prd{Phys.~Rev.~D}%
 \newcommand\be{\begin{equation}}
 \newcommand\ee{\end{equation}}
 \def\bib#1#2#3#4#5#6#7#8{\bibitem{#1} {#2}\ {#3}, {{#5}},\  {{#6}},\  {#7}  \par }

\begin{document}

 \title{INPOP08, a 4-D planetary ephemeris: \\
  From asteroid and time-scale computations to ESA Mars Express and Venus Express contributions.}

 \author{A. Fienga\inst{1,2}
 \and J. Laskar\inst{1}
 \and T. Morley \inst{3}
 \and H. Manche\inst{1}
 \and P. Kuchynka\inst{1}
 \and C. Le Poncin-Lafitte\inst{4} 
 \and F. Budnik \inst{3}
 \and M. Gastineau\inst{1}
 \and L. Somenzi \inst{1,2}
 }

 %\offprints{A. Fienga}
 \institute{Astronomie et Syst\`emes Dynamiques,
 IMCCE-CNRS UMR8028,
 77 Av. Denfert-Rochereau, 75014 Paris, France
 \and
 Observatoire de Besan\c con, CNRS UMR6213,
 41bis Av. de l'Observatoire, 25000 Besan\c con, France
 \and
ESOC, Robert-Bosch-Str. 5, Darmstadt, D-64293 Germany
 \and
 SYRTE, CNRS UMR8630,
 Observatoire de Paris,
 77 Av. Denfert-Rochereau, 75014 Paris, France
 }

 \offprints{A. Fienga, fienga@imcce.fr}

 \date{\today}

 \titlerunning{INPOP08, a 4-D planetary ephemeris}
 \authorrunning{Fienga et al}

  \abstract{
The latest version of the planetary ephemerides developed at the Paris Observatory and at the Besan\c con Observatory is presented here.
INPOP08 is a 4-dimension ephemeris since it provides to users positions and velocities of planets and the relation between TT and TDB.
Investigations leading to improve the modeling of asteroids are described as well as the new sets of observations used for the fit of INPOP08.
New observations provided by the European Space Agency (ESA) deduced from the tracking of the Mars Express (MEX) and Venus Express (VEX) missions are presented as well as the normal point deduced from the Cassini mission. We show the huge impact brought by these observations in the fit of INPOP08, especially in terms of Venus, Saturn and Earth-Moon barycenter orbits.
 }
 \maketitle
 %
 %______________

 \section{Introduction}

 Since the first release, INPOP06,
 of the  planetary ephemerides developed 
 at Paris and Besan\c con Observatories,
 (Fienga et al. 2008, www.imcce.fr/inpop), 
 several  improvements  have been conducted on the  dynamical modeling
 of the INPOP ephemeris. 
The observation dataset has also been substantially increased, 
especially  
 with the addition of ranging data from the ESA space missions Mars Express and Venus Express. 
%Moreover, 
% the perspective of highly accurate astrometric space missions like GAIA {\bf{ or highly accurate pulsar timing astrometry}}, 
% led to increase the consistency of the ephemeris reference frames 
% by adopting time-scales that are in agreement with the definitions adopted  
% by the International Astronomical Union in 2006. 
The resulting new version of INPOP planetary ephemerides, INPOP08, is presented here 
 with the description of its new features.
 The INPOP08 ephemeris has also been fitted to all available 
 Lunar Laser Ranging data (Manche et al., 2007).
 
 One of the novelties present in INPOP08 is the addition, in the distribution of the ephemeris, of a time scale transformation TT-TDB 
 that is coherent with the ephemeris.
%Indeed, taking into account that the relativistic modeling of ongoing space astrometry missions, such as GAIA, needs ephemerides fully consistent with the basic principles of General Relativity, we propose a new procedure with INPOP08, enabling us to go towards four-dimensional planetary ephemerides. 
The basic idea is to provide to users positions and velocities of Solar System celestial objects, and also, time ephemerides relating the Terrestrial time-scale, TT, and the time argument of INPOP, the so-called Barycentric Dynamical Time TDB, based on the definition adopted by the International Astronomical Union in 2006.
Such a release of planet and time ephemerides enable us to go towards four-dimensional planetary ephemerides.
Section \ref{TTTDB} describes the   INPOP procedure 
 for the computation of the TT-TDB relation.
  
%As this  TT-TDB relation is dependent on each planetary ephemeris, 
%  from now on, the INPOP deliveries will provide to users positions, 
%  velocities of planets and TT-TDB. This will garantee to the INPOP 
% ephemerides a complete consistency in terms of relativistic metrics.

 Section \ref{ringtxt} is devoted to a brief account of  the new constraints and modeling implemented in INPOP for the asteroid perturbations. 
 As in INPOP06,  300 asteroids are  included in the dynamical equations of INPOP08, and  the remaining ones are modelized as a ring. With respect to INPOP06, the ring model and the asteroid selection have been 
 improved, and the precise description of these 
 advances are made in (Kuchynka et al, 2009).

 An important part of the new observations used for the INPOP08 fit consists in the tracking  data provided  by the   ESA space missions Mars Express and Venus Express (Morley 2006a, 2007a, 2007b). 
These datasets are the first  radio ranging data provided by ESA, 
 and their acquisition and reduction process is described in section 
 \ref{mexvexdata}.
 VEX observations bring a very important set of informations on Venus orbit. It is especially of interest since this orbit is much less perturbed by asteroids than Mars and therefore has greater potential for precise ephemeris motion, fundamental physics testing and reference frame establishment.

  The section 5 deals with the INPOP fit obtained by comparisons between the dynamical modeling and the  observations. In this process, 34 asteroid 
  masses were fitted against only 5 in INPOP06. 
 Comparisons are made between obtained asteroid masses and other published masses. 
Values for the fitted Earth-Moon barycenter and Sun oblateness $J_2$ are also given.
We have also fitted the AU and comparisons are provided with the latest determinations of 
 DE414 (Standish  2006; Konopliv et al. 2006) and DE421 (Folkner  2008).
 %In addition, {\bf {as we believe that the AU should be assigned to a fixed value, 
 %we have performed as well a second fit, where the AU is given the IERS Conventions 2003 value (IERS03), and the Solar GM is fitted. 
 
In addition, we have performed as well a second fit, where the AU is given the IERS Conventions 2003 value (IERS03), and the Solar GM is deduced.
 
On several occasions, for planned high precision observations, 
we had some enquiries about the real accuracy of the 
position or velocities given by the planetary ephemerides. This is 
actually a difficult  question to answer, but we have tried in section 
\ref{uncert} to provide some estimates of these uncertainties, by comparison
with INPOP06 and DE421. The last section is devoted to the conclusions and perspectives. 

 \section{INPOP as a 4-D planetary ephemeris}
\label{TTTDB}

 With INPOP08, we aim to produce planetary ephemerides as fully compatible as possible with the relativistic background recently adopted by the 
 astronomical community and summarized by the IAU2000 and IAU2006 conventions
 (Soffel el al., 2003). This leads to the production of an ephemeris in TDB, and  to the construction of a TT-TDB transformation. The following subsections describe
 the various steps involved in this process and the impact 
 for the ephemeris users.

 \subsection{The TCB-TCG transformation}
 Two reference systems are defined: a global one, the Barycentric Celestial Reference System (BCRS), covering the whole Solar System and a local one, the Geocentric Celestial Reference System (GCRS), which is physically suitable for the modeling of processes in the vicinity of the Earth.
BCRS is particularly useful when one wants to model the light propagation or motion of celestial objects in the Solar System. We can then, in the mass-monopoles approximation, write the equation of motion of bodies as well as the conservation laws satisfied by the Solar System barycenter (see Damour and Vokrouhlick{\'y} 1995); all these features being already implemented in INPOP06. (Damour and Vokrouhlick{\'y} 1995) gives the complete conservation equations based on (Damour,Soffel and Xu 1991) formalism including spin/spin, mass/mass and mass/spin couplings.
However, each fundamental reference system has its own time-scale: TCB for the BCRS and TCG for the GCRS. The relation between TCB and TCG can be derived, but  time transformations are only determined for specified space-time events (one coordinate time and three spatial positions) so, at the geocenter, the transformation reads (Damour, Soffel and Xu 1991, Soffel et al. 2003):
 %\begin{eqnarray}
 %\begin{eqnarray}
 %\label{BCRS1}g_{00}&=&-1+\frac{2}{c2}w(t,\vec{x})-\frac{2}{c4}w2(t,\vec{x})+\mathcal{O}\left(\frac{1}{c6}\right)\, ,\\
 %g_{0i}&=&-\frac{4}{c3}w^i(t,\vec{x})+\mathcal{O}\left(\frac{1}{c5}\right)\, , \\
 %g_{ij}&=&\delta_{ij}\left\lbrack 1+\frac{2}{c2}w(t,\vec{x})\right\rbrack+\mathcal{O}\left(\frac{1}{c4}\right)\, ,
 %\end{eqnarray}
 %with $c$ is the velocity of light in a vacuum, the potentials $w$ and $w^i$ being given in the mass-monopoles approximation by
 %\begin{eqnarray}
 %w(t,\vec{x})&=&\sum_A\frac{GM_A}{r_A}+\frac{1}{c2}\sum_A\frac{GM_A}{r_A}\Bigg\lbrack 2v_A2-\sum_{B\ne A}\frac{GM_B}{r_{BA}}\nonumber\\
 %&&\qquad\qquad\left.-\frac{1}{2}\left(\frac{(\vec{r}_A.\vec{v}_A)^2}{r_A2}+\vec{r}_A.\vec{a}_A\right)\right\rbrack\, ,\\
 %\label{BCRS2}w^i(t,\vec{x})&=&\sum_A\frac{GM_A}{r_A}v_A^i\, ,
 %\end{eqnarray}
 %where capital latin subscripts $A$, $B$ enumerate massive bodies, $M_A$ is the mass of body $A$, $\vec{r}_{BA}=\vec{x}_B-\vec{x}_A$, $r_{BA}=\vert\vec{r}_{BA}\vert$, $\vec{v}_A=\dot{\vec{x}}_A$, $\vec{a}_A=\dot{\vec{v}}_A$, a dot signifying time derivative with respect to TCB and $\vec{x}_A$ being the BCRS position of body $A$.
 \begin{equation}\label{TCB2TCG}
\frac{dTCG}{dTCB}=1+\frac{1}{c^2}\alpha(TCB)+\frac{1}{c^4}\beta(TCB)+\mathcal{O}\left(\frac{1}{c^5}\right)
 \end{equation}
 $c$ being the speed of light in a vacuum and where
 \begin{eqnarray}
 \alpha(TCB)&=&-\frac{1}{2}v_E^2-\sum_{A\ne E}\frac{GM_A}{r_{EA}}\, ,\label{TCB2TCG1}\\
 \beta(TCB)&=&-\frac{1}{8}v_E^4+\frac{1}{2}\left\lbrack\sum_{A\ne E}\frac{GM_A}{r_{EA}}\right\rbrack^2\nonumber\\
 &&+\sum_{A\ne E}\frac{GM_A}{r_{EA}}\Biggl\lbrace 4\vec{v}_A.\vec{v}_E-\frac{3}{2}v_E^2-2v_A^2\nonumber\\
 &&\qquad\qquad \qquad +\frac{1}{2}\vec{a}_A.\vec{r}_{EA}+\frac{1}{2}\left(\frac{\vec{v}_A.\vec{r}_{EA}}{r_{EA}}\right)^2 \nonumber\\
 &&\qquad\qquad \qquad+\sum_{B\ne A}\frac{GM_B}{r_{AB}}\Biggr\rbrace\, . \label{TCB2TCG2}
 \end{eqnarray}
 Capital latin subscripts $A$ and $B$ enumerate massive bodies, $E$ corresponds to the Earth, $M_A$ is the mass of body $A$, $\vec{r}_{EA}=\vec{x}_E-\vec{x}_A$, $r_{EA}=\vert\vec{r}_{EA}\vert$, $\vec{v}_A=\dot{\vec{x}}_A$, $\vec{a}_A=\dot{\vec{v}}_A$, a dot signifying time derivative with respect to TCB and $\vec{x}_A$ being the BCRS position of body $A$. 

 Eqs. (\ref{TCB2TCG}), (\ref{TCB2TCG1}) and (\ref{TCB2TCG2}) exhibit that the time transformation between TCB and TCG is  explicitly related to the positions and velocities of Solar System bodies, so to the planetary ephemeris itself. As stressed by Klioner (2008), if we are distributing 
 a time transformation together with
positions and velocities, we are building a four-dimensional planetary ephemeris. 
This time transformation will be called in the following time ephemeris to be close to the terminology of Irwin and Fukushima (1999). %This is exactly the choice done for INPOP08, taking into account that 4D ephemerides will become necessary for the relativistic modeling of the ongoing generation of space astrometry missions like GAIA.

An issue remains. The use of TCB in planetary ephemerides should induce important changes in numerical values of planet masses, initial conditions and astronomical unit commonly adopted by users.

 Therefore, even if a TCB-based ephemeris and a TDB-based ephemeris are related linearly, caution has to be taken when one wants to shift from one ephemeris to an other as it will induce large changes for all users. For a more complete discussion, see for instance (Klioner 2008).
 
 \subsection{The TT-TCG transformation}
 The IAU time realization is done by the International Atomic Time (TAI) with a service running since 1958 and attempting to match the rate of proper time on the geoid by using an ensemble of atomic clocks spread over the surface and low orbital space of the Earth. TAI is conventionally related to the Terrestrial Time (TT) and the Universal Coordinate Time (UTC) by the two following relations:
 \begin{equation}
 \label{TT2TAI}
 TT(TAI)=TAI+32.184s\, .
 \end{equation}
 and
 \begin{equation}
 \label{UTC2TAI}
 UTC=TAI-\mbox{leap seconds}\, ,
 \end{equation}
 leap seconds being added by the International Earth Rotation Service at irregular intervals to compensate for the Earth's rotation irregularities. Mainly, leap seconds are used to allow UTC to closely track UT1, which effectively represents the Earth rotation.\newline
It is thus more convenient to consider a regular time-scale like TT. Moreover due to its IAU1991 definition, TT is related to TCG by a fixed linear function as follows:
 \begin{equation}
 \label{TT2TCG}
 TCG-TT=L_G \times (JD-2443144.5)\times 86400 \, ,
 \end{equation}
 where $L_G=6.969290134\times 10^{-10}$ and JD is TAI measured in Julian days.

\subsection{The TDB-TCB transformation}
 
The TDB has a more tumultuous history. It was introduced by IAU1976 in order to remain close to TT up to periodic variations. However such a definition was flawed because in that case TDB cannot be a linear function of TCB. 
Consequently,  the relativistic equations 
derived in TCB cannot be simply adapted to TDB. 
In practice, Caltech/JPL, Harvard/CfA, and RAS/IAA ephemeris programs have used a relativistic coordinate time whose
mean rate is automatically adjusted to the mean rate of TT by way
of the ephemeris fits. As such, those times, often referred to as
T$_{eph}$, can be related to TCB by the following equation with the notation of equation \ref{TCB2TCG}:
\begin{equation}
 \label{Teph}
 \frac{dT_{eph}}{TCB} = 1 + \frac{1}{c^{2}} \alpha(TCB)
 \end{equation}
and then from (Irwin and Fukushima 1999), 
\begin{equation}
TCB \equiv \frac{1}{1-L_{B}} (T_{eph} - T_{eph0}).
\end{equation}
Differences between T$_{eph}$ and TCB have been estimated by numerical quadrature method of equation \ref{Teph} which is equivalent to equation \ref{dTTmTDB_dTDB} limited at the $1/c^{2}$ terms. 
Chebychev polynomials of the solutions were provided to users and corrections were brought to the analytical representation of TDB done few years earlier by Fairhead and Bretagnon (1990). 

This situation changed  with the recommendation B3 of IAU2006 which made TDB a fixed linear function of TCB as follows:
 \begin{equation}
 \label{TDB2TCB}
 TDB = TCB-L_B(JD-T_0 )\times 86400+TDB_0,
 \end{equation}
 where $T_0 = 2443144.5003725$, $L_B=1.550519768\times 10^{-8}$, $TDB_0=-6.55\times10^{-5}$s and  $JD$ is the TCB Julian date. TCB value (like TT and TCG ones) is $T_0$ for the event 1977 January 1 00h 00m 00s TAI at the geocenter. It increases by one for each 86400s of TCB.
%This TDB definition is given as an improvement (IAU2006 conventions) in the consistency of the equations of motion in the General Relativity background with a relative of continuity with the former TDB .
 The use of a TDB planetary ephemeris as defined in equation (\ref{TDB2TCB}) has no significant impact for normal users: values of masses and initial conditions of solar system objects being the same at the common level of accuracy.

%By shifting from the previous TDB time-scale as described by Standish (1998a) to the the newly defined TDB, INPOP08 makes a compromise between the consistency of the basic General Relativity metrics and the users convenience.

 \subsection{Implementation in INPOP08}
From equations (\ref{TCB2TCG}), (\ref{TT2TCG}) and (\ref{TDB2TCB}), we have (Klioner, 2007):

\begin{eqnarray}\label{dTTmTDB_dTDB}
 \frac{d\left(TT-TDB   \right)}{dTDB} & = & \left( L_B + \frac{1}{c^2}\alpha \right) \left(1+L_B-L_G \right) -L_G \nonumber \\ 
                                                         &    &  + \frac{1}{c^4}\beta
\end{eqnarray}

One can notice that the values of $\alpha$ and $\beta$ do not change when using quantities (GM, positions, velocities and accelerations of bodies) expressed in TDB instead of TCB units.
It is then straightforward to construct a time ephemeris $TT-TDB = f(TDB)$ (noted $\Delta TDB=f(TDB)$).
%TT$\leftrightarrow$TDB
This is the choice done for INPOP08.

The central part of our implementation into INPOP is to numerically integrate Eq. (\ref{dTTmTDB_dTDB}) together with the equations of motions of all bodies. 
Because the right member of (\ref{dTTmTDB_dTDB}) does not depend on $\Delta TDB$, the equation (\ref{dTTmTDB_dTDB}) is not strictly an ordinary differential equation.
In the computation of $\alpha$, $A$ enumerates all massive bodies of the Solar System, that is,  the Sun, the planets, Pluto, the Moon and all the 303 asteroids.
In $\beta$ (which is divided by $c^4$, and so is less important than $\alpha$), $A$ and $B$ enumerate all bodies except the 298 ``small'' asteroids. 
For the acceleration term $\vec {a}_A$, all the interactions are taken into account, including newtonian interactions, relativistic corrections, figure and tide effects. It was already needed for the equation of motion of the corresponding body, and no additional work is therefore necessary to compute it.

Because the $(TT-TDB)$ value is unknown at J2000 (it depends on the ephemeris), the initial condition is set to zero. The quantity integrated in the state vector (including positions and velocities vectors of bodies) is then $\Delta TDB + k$ where $k$ is an offset determined later, just before building the Chebychev polynomials, by using the following condition: for the event 1977 January 1st 00h 00m 00s TAI at the geocenter, TDB julian day is $T_0+TDB_0$ and $\Delta TDB = -TDB_0$.

At this point, the difference $TT-TDB$ can be computed from any value of $TDB$. But in the reduction process of observations and because they are dated in UTC time scale, the transformation $TT-TDB$ as a function of  $TT$ (noted $\Delta TT$) is needed.
 
 A similar differential equation as eq. (\ref{dTTmTDB_dTDB}) can be found for $d(\Delta TT)/dTT$ (Klioner, 2007), but it is not necessary to integrate it.
From the relation $(TT-TDB)=f(TDB)$ computed with INPOP, one can notice that $(TT-TDB)=f(TT-(TT-TDB))$. $\Delta TT$ is then solution of an implicit equation, which can be solved by iterations. In fact, only one is necessary, discrepancies between $TT$ and $TDB$ being smaller than 2 milliseconds.

\begin{figure}
 \includegraphics[width=\figw]{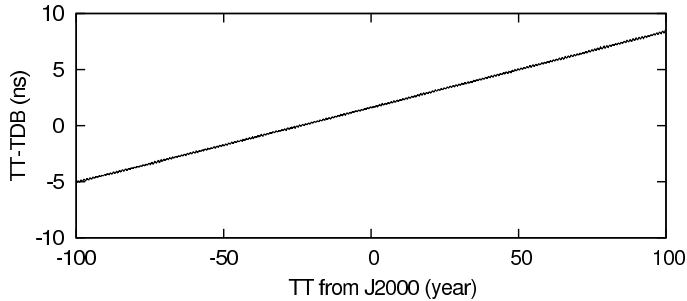}
 \caption{Differences (in nanosecond) between TE405 (corrected from the $6.55 \times 10^{-5}$ second offset) and the (TT-TDB) integrated in INPOP08. X axis is Terrestrial Time, expressed in years from J2000.}
 \label{figTTmTDB}
 \end{figure}

The formal differences between the present procedure and TE405 (Irwin \& Fukushima, 1999) are small as illustrated in Fig.~\ref{figTTmTDB}. The linear drift is measured to $6.74$ nanosecond per century ($\mbox{ns/cy}$). This value is consistent with the IAU2006 resolution B3 (less than 1 nanosecond per year between TT and TDB). These small discrepancies could be surprising because TE405 does not take into account the terms in $1/c^4$. Neglecting them in eq. (\ref{dTTmTDB_dTDB})  induces an important drift of $346.0 \mbox{ ns/cy}$ , close to the value $\Delta L_C^{(PN)}= 346.2 \mbox{ ns/cy}$ from (Irwin \& Fukushima, 1999). But this drift can be compensated by a change of the constant $L_B$ or $L_C$. Now that TDB is defined as a conventional and fixed linear function of TCB (see eq. \ref{TDB2TCB}) and $L_B$ is fixed to the value given by the IAU 2006 resolution B3, terms in $1/c^4$ are thus essential.
 
Because  discrepancies are small,  no impact is noticeable in the differences between observed planet positions and positions computed with TE405, the relation of Fairhead and Bretagnon (1990), or with the Chebychev polynomials representing the INPOP TT$\leftrightarrow$TDB. No additional iteration is then necessary, even if at each step of the adjustment of INPOP to observations, the computation of TT$\leftrightarrow$TDB coefficients is automatically iterated.

 %{\bf Nota Bene : Pour avoir une ephemeride 4D, il faut fournir les deux fonctions en Chebyshev ainsi que 2 routines pour les lire : TT2TDB et TDB2TT. L'input de TT2TDB doit etre un temps en TT et calcule le TDB correspondant. A l'inverse TDB2TT a pour input un temps en TDB et calcule le TT correspondant. Ajout de Commentaire: en realite, une seule relation suffit. }\newline

\section{New constraint on asteroid modeling}
\label{ringtxt}

\subsection{Supplementary selection of asteroids}
For INPOP08, we also slightly revised the INPOP06 selection of asteroids perturbing Mars and the inner planets. 
This was done with the same method as the estimation of the mass of the ring described in the \ref{ring} section : 24635 selected asteroids are assigned with reasonable distribution of masses according to available data and the Statistical Asteroid Model (Tedesco et al. 2003). 
A Monte Carlo study allows to assign to each asteroid the probability of being among the 300 most perturbing asteroids in terms of amplitude of the perturbation on the Earth-Mars distance between 1969 and 2010: 
For each of the 24635 asteroids, two integrations have been done between 1969 and 2010: one with the asteroid $i$ and one without the asteroid $i$, $i$ varying from 1 to 24635. The differences between the two integrations give the impact induced by the asteroid $i$. To built our list of asteroids, we have studied the impact not only on the Earth-Mars distances, but also the Earth-Venus and Earth-Mercury distances.
We are thus able to compile the most probable list of the 300 most perturbing asteroids and compare it to the INPOP06 selection used in INPOP06\footnote[1]{the list of asteroids integrated individually in INPOP06 or DE405 is given in (Standish and Williams 2001)}. 
Three asteroids have predicted perturbation amplitudes well over 30m and are absent from the INPOP06 selection: 60 Echo (amplitude reaches 170m), 585 Bilkis and 516 Amherstia.
We added these to the 300 already integrated in INPOP06 and thus there are in total 303 asteroids integrated individually in INPOP08. 
More details on the methodology used for obtaining the most probable list of the top 300 perturbing asteroids can be found in (Kuchynka et al. 2009).

 \subsection{The ring}
 \label{ring}

In the former version of INPOP (Fienga et al. 2008), perturbations of asteroids on planets were modeled with 300 individual asteroids and a static circular ring at 2.8 AU. Five asteroid masses, 3 taxonomic densities (attributed to the remaining 295 asteroids) and the mass of the ring were fitted to observations.

After the integration of such a model on a 100-years time interval, it appeared that the asteroid ring induced a drift of several meters in the position of the Solar System barycenter (Fig.\ref{ring}). The static ring was then replaced by a more realistic implementation that now conserves the total linear and angular momenta of the system: the ring interacts fully with the planets and is no longer assumed to act only in their ecliptic planes. Its center is attached to the Sun and its orientation is an integrated parameter which evolves with time. The ring's interaction with other objects is actually identical to the interactions of an asteroid on a circular orbit averaged over the asteroid's mean orbital motion. Besides eliminating the barycenter drift, the advantage of the new implementation is that the ring is taken into account in a more realistic way and its presence in the model is more meaningful. The ring's radius was chosen at 3.14 AU.

Another novelty is that the mass of the ring is not fitted to observations but estimated independently. 
This estimation is made by calculating the amplitude of the perturbation on the Earth-Mars distance exerted by all main-belt asteroids but the 303 most perturbing ones determined previously. 
24635 asteroids are considered as a model of the main belt and a simple scheme based on the Statistical Asteroid Model (Tedesco et al. 2005) is used to assign each asteroid with a reasonable distribution of masses. 
A Monte Carlo experiment where asteroids are assigned random (but reasonable) masses allows to calculate the corresponding perturbation from all the asteroids but the 303 most perturbing ones. 
For each random set of masses, a ring's mass is determined so as to fit the perturbation of the ring to the global effect. 
This leads to the estimation of the ring's mass at $M_{ring} = (1\pm0.3) \times 10^{-10} M_{\bigodot}$ for a ring at 3.14 AU (Kuchynka et al. 2009).

 \begin{figure}
 \includegraphics[width=\figw]{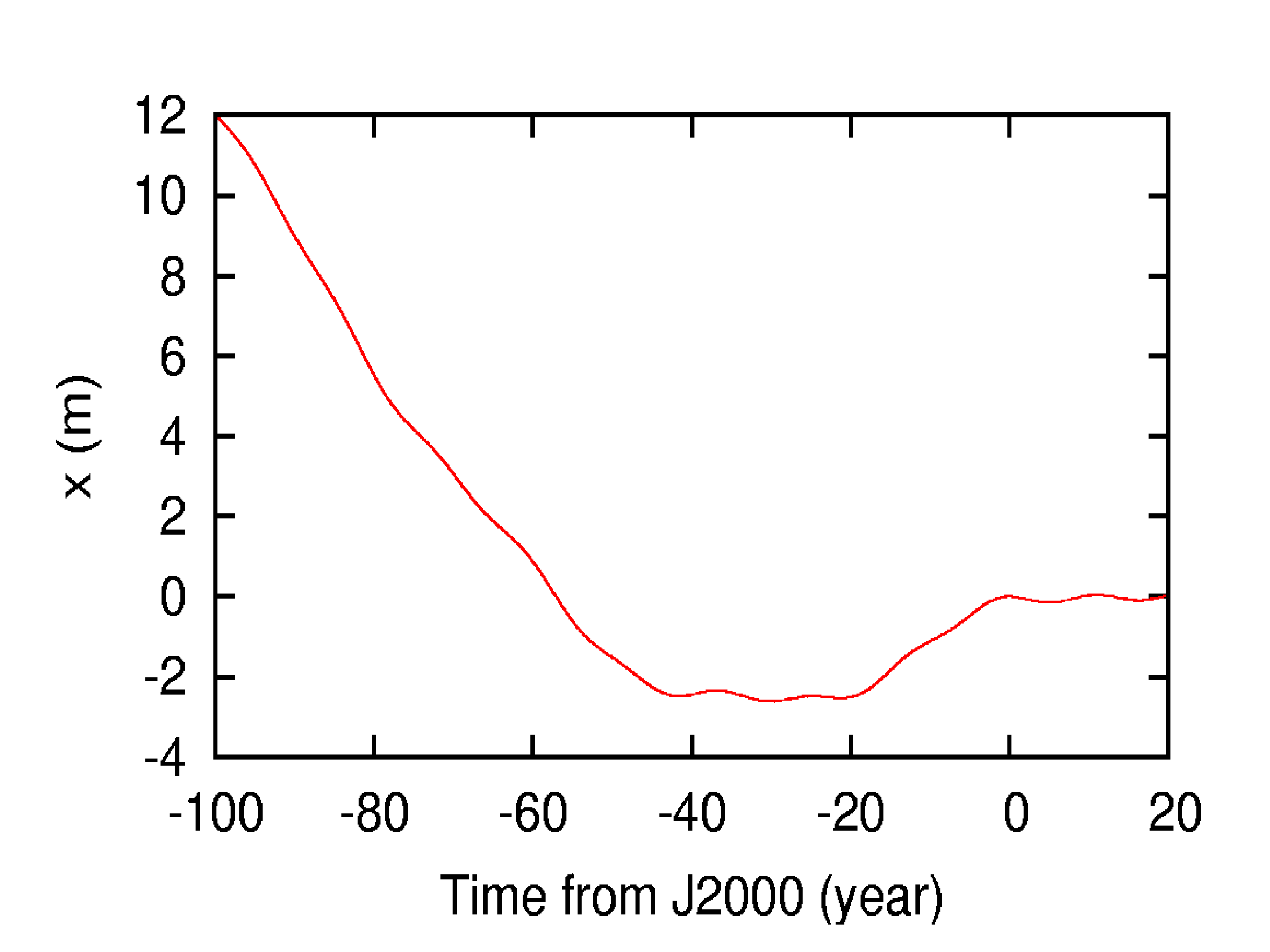}
 \caption{Impact of the asteroid ring on the position of the SSB over 100-years integration.}
 \label{ring}
 \end{figure}

 \section{Presentation of MEX and VEX observations}
 \label{mexvexdata}
 The tracking of Mars Express (MEX) and Venus Express (VEX), while orbiting their
 planetary namesakes, comprises two-way coherent Doppler and range measurements. The
 Doppler data are converted to range-rate: the component of the spacecraft's velocity
 relative to the ground station along the direction from the station to the spacecraft. The determination of the spacecraft orbits uses only the range-rate data since, under normal
 circumstances, the additional inclusion of the range data leads to only insignificant
 improvement in the accuracy of the orbit solutions.
 
 Within the orbit determination, the computed values of the observables rely on high fidelity modeling of the dynamics and the signal path (Budnik et al. 2004; Moyer 2000). As
 part of this modeling, the orbital states of Mars and Venus are taken from the JPL DE405
 planetary ephemerides (Standish 1998b).
 Over time the range residuals exhibit a signature whose predominant contribution is the error in the distance from the Earth to the planet as given by the planetary ephemerides.
 The range residuals are therefore derived data that are very useful for 
 improving the accuracy of planetary ephemerides.
 The accuracy of the range residuals depends on a number of factors, which are described in the following subsections.

 \subsection{Accuracy of the MEX orbit determinations}
 %\subsubsection{Mars Express}
 MEX, the first European mission to Mars, was launched on 2nd June 2003 and inserted
 into Mars orbit on 25 December 2003 and is presently extended until May 2009.

 MEX is tracked with both ESA 35 m deep space antennas and those of the NASA Deep
 Space Network (DSN) (mainly at the Goldstone complex), in almost equal proportions in
 terms of tracking duration. Most of the ESA tracking is from New Norcia
 (NNO), in Western Australia, the rest from Cebreros, near Madrid in Spain.

 The quantity of DSN range data and hence
 range residuals is far higher than the quantity of ESA range residuals due to the extraction of one ESA raw data point every 20 minutes whereas DSN range measurements of MEX are made at intervals of 3 minutes 27 seconds.
 The standard deviation of range residuals for individual passes gives a good indication of
 the random noise on the measurements. For ESA data, the average standard deviation of
 the (two-way) residuals is less than 1 m. For NASA/DSN data it is below 0.5 m.

 In terms of orbit geometry, the operational orbit is near polar and elliptic with an apoapsis altitude of a little over 10000 km and a periapsis altitude that has varied between 250 km and 340 km. The orbital
period averaged 6.72 hours until late in 2007 when a series of five manoeuvres at periapsis
increased the period to 6.84 hours.
 MEX is equipped with a fixed high-gain antenna (HGA) that must be pointed towards the
 Earth during tracking passes and tracking is done in X-band
up- and downlink. The primary science data collection is around periapsis passage when the spacecraft HGA is not Earth-pointing. For orbit determination purposes, the information content within the Doppler data is highest around periapsis,
 so the orbit solution accuracy is adversely affected by the lack of such data. The orbit
 determination accuracy is also limited by other factors; most notably the imperfect calibration of perturbing velocity increments caused by thrusting, to off-load the accumulated angular momentum of
 the reaction wheels, and also the problems in accurately modeling small forces due to
 solar radiation pressure and the tiny but highly variable atmospheric drag at each periapsis
 passage.

 For much of the MEX operational mission, routine orbit determinations were made twice
 per week based on tracking data arcs of 5-7 days duration, corresponding to approximately
 18-25 orbital revolutions, with a typical overlap of 2 days between successive arcs. Nowadays, orbit determination is made weekly using an arc of about 10 days duration and therefore with similar overlaps. The differences between the values of common range residuals
 computed from successive orbit determinations provide an indication of the effect of
 orbit determination errors on the accuracy of the residuals. Outside of periods of superior
 solar conjunction, these differences are almost always below 3 m. The
 range residuals that are retained are taken from the earlier of each overlapping solution.

 %\subsubsection{MEX range residuals}
 %Figure 1 is a plot of the 76164 MEX %two-way range residuals from 5th March 2005 until
 %1st April 2008. Ignoring the interval around the superior solar conjunction in autumn
 %2006, the signature shows that the error in the predicted value from the DE405 ephemeri-
 %des of the Earth-Mars distance has varied between -200 m and +170 m.
 %During September 2006, MEX experienced long eclipses and to conserve power the on-
 %oard transmitter was switched on only occasionally. Very few tracking data were
 %acquired and when ranging measurements were available they were used together with the
 %Doppler data for orbit determination. This explains the first data gap in Figure 1. The sec-
 %ond gap is between 8th November and 7th December 2006. At the request of the Mars
 %Global Surveyor (MGS) project, MEX ranging was not performed so as to guarantee no
 %interference with communications to MGS that had been lost (and which were never
 %recovered).

 %\subsection{Venus Express}
 \subsection{Accuracy of the VEX orbit determinations}
 VEX, the first European mission to Venus, was launched on 9th November 2005 and
 inserted into Venus orbit on 11th April 2006. Its mission is presently extended until May
 2009.

 The primary ground station supporting VEX is Cebreros, which tracks the spacecraft
 almost every day. Between 27th April 2006 and 13th March 2008, there were 663 Cebreros passes whose accumulated duration accounts for 97.7 \% of the total tracking duration.
 New Norcia has provided two-way coherent radiometric data during 12 passes and NASA
 DSN stations during 26 passes, of which 16 were made with DSS 43 at Canberra, mainly
 during the superior solar conjunction in autumn 2006.
 The sampling rates of the range data and the random noise on the measurements are the
 same as for MEX.

 The operational orbit is polar and highly elliptic. There is hardly any precession of
 the line of apsides, so the periapsis, whose argument is currently at 95$^\circ$, remains close to
 the planet north pole. The apoapsis altitude is about 66500 km and the periapsis altitude
 has been controlled to stay within the range of 185 km to 390 km. The orbital period is
 nominally 24 hours and the maximum excursion has never been more than 6 minutes.
 Every orbital revolution, in the vicinity of apoapsis but not within ground station visibility, the momentum of the reaction wheels is off-loaded by thrusting. The perturbing $\Delta V$
 into the orbit is typically in the range 15-25 mm/s which is substantially higher than for
 the MEX wheel off-loadings. The spacecraft attitude and the direction of the thrust are
 chosen so that the manoeuvres help to control the orbit phasing. The control is such that
 the signal elevation from the daily Cebreros passes rises to 10 $^\circ$ (when telecommanding
 may start) always close to 2 hours after periapsis passage. The pass ends either 10 hours
 later, close to apoapsis, or at 10$^\circ$ descending elevation, whichever is earlier. Tracking
 data obtained in X-band up and downlink are thus almost never acquired during the descending leg of the orbit, nor around periapsis.
 The combination of the unfavourable pattern of tracking data arcs, imperfect calibration of
 the wheel off-loadings and deficiencies in modeling forces due to solar radiation pressure,
 together with other factors depending on the nature of the orbit cause the accuracy of
 the orbit determination to be worse than that for MEX.
  Typical values of the differences between range residuals derived from successive
 orbit solutions are a few metres but, occasionally, even away from solar conjunction periods, the differences can reach as high as 10 m.

 %\subsubsection{VEX range residuals}
 %Figure 2 is a plot of the 18668 VEX two-way range residuals from 27th April 2006 until
 %20th April 2008. The signature shows that the error in the predicted value from the DE405
 %ephemerides of the Earth-Venus distance has varied between -160 m and +280 m.
 %1. In August 2008 it is planned to lower the periapsis altitude to a minimum of 185 km as a preliminary
 %operation for further investigations into the tenuous upper atmosphere of Venus.

 %\subsection{Accuracy of the range residuals}
 %Additional to orbit determination accuracy, there are other factors which affect the accuracy of the range residuals.

 \subsection{Spacecraft transponder group delay}
 Subtracted from each range measurement is the nominal value of the group delay of the
 on-board transponder. For MEX, the value corresponding to the normally used X-band
 up- and downlink signals is 2076 nanoseconds (about 622 m). For VEX, that has a virtually
 identical transponder, the value is 2085 ns (about 625 m).
 The nominal value is the average value measured at different occasions on ground before
 launch. From the variations in measured values, ostensibly made under identical conditions, it is thought that the systematic
 error of the nominal value should not be larger than 30 ns (about 10 m). In addition, it is known
 that the group delay is not perfectly stable and can fluctuate by a few ns, depending upon
 variations in a number of parameters such as temperature and signal strength.
 For MEX, these error estimates appear reasonable, and perhaps a little conservative, based
 upon the consistency with results determined from the NASA MGS and MO spacecraft range data
 during the spring of 2005. For VEX, there is no independent means to verify the error estimates.

 \subsection{Superior solar conjunction}
\begin{figure*}
 \begin{center}
 \includegraphics[width=12cm]{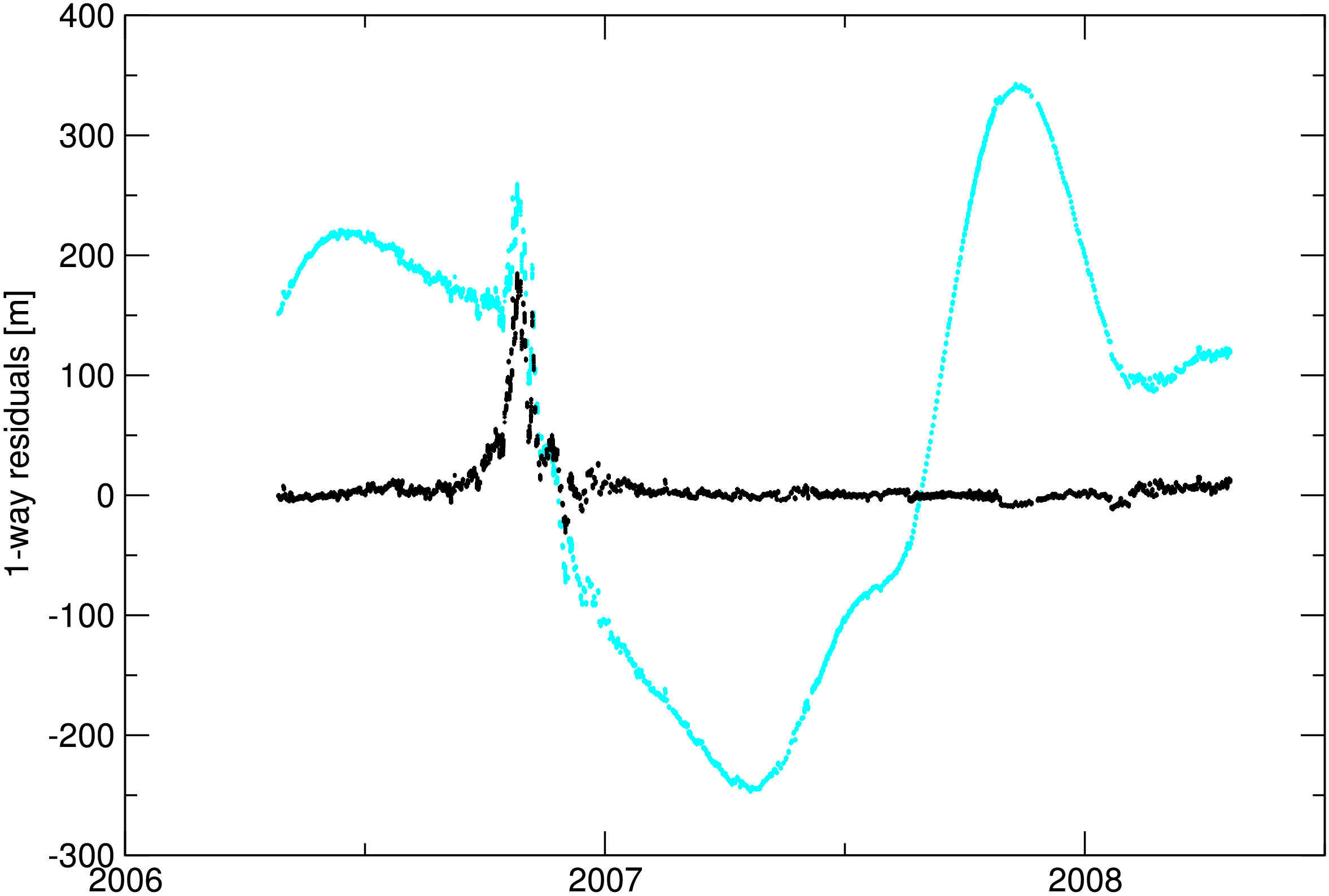}
 \caption{VEX 1-way residuals with the new INPOP08 fitted to VEX and MEX data (dark curve) and INPOP06 not fitted on VEX observations (light curve). The solar conjonction is clearly identified with the peak of residuals in october 2006.}
 \label{vexplot}
 \end{center}
 \end{figure*}

 During the time that MEX and VEX range residuals have been generated and archived,
 both missions have experienced a superior solar conjunction. The MEX Sun-Earth-probe
 (SEP) angle remained below 10$^\circ$ for two months centred on 23rd October 2006, when the
 minimum SEP angle was 0.39$^\circ$ (1.6 solar radii). The VEX SEP angle was continuously
 less than 8$^\circ$ over two months centred on 27th October 2006, when the minimum SEP
 angle was 0.95$^\circ$ (3.8 solar radii).
 The effects on spacecraft radiometric data at these conjunctions have been described by
 (Morley and Budnik, 2007). The signals to and from the spacecraft pass through the solar
 corona surrounding the Sun. The free electrons in the plasma cause a group delay on ranging measurements.
 Since the electron density increases with decreasing distance from the
 Sun, following, at least approximately, an inverse square law, the delay increases as the
 SEP angle diminishes.
 No solar corona model was applied when computing the range residuals, so the increased
 delay is the cause of the peak in figure \ref{vexplot} and gaps in figure \ref{mexplot}.
 The existing solar corona models that could be used to correct the range residuals are not
 very accurate. They cannot take into account the quite large day-to-day variations in the
 signal delay caused by short-term fluctuations in solar activity like sunspot formation,
 flares and coronal mass ejections, all of which can influence the surrounding electron density.
 A secondary cause of increased errors in range residuals at solar conjunction is due to the
 main effect on Doppler measurements of a substantial increase in noise. When the SEP
 angle falls to about 1$^\circ$ , the measurement noise typically increases up to two orders of
 magnitude higher than is usual at large SEP angles. The accuracy of range residuals is
 then indirectly and adversely affected by a degradation in the accuracy of the orbit determination solutions.
 That is why the decision was taken to omit 30 days before and after the solar conjunction in range residuals.
 To illustrate the impact of solar conjunction, peak in residuals is plotted in the case of VEX data on figure \ref{vexplot}. On figure \ref{mexplot}, the gap in MEX data around October 2006 is induced by the solar conjunction. Such solar conjunctions  also occured during the MGS and MO missions and caused observation gaps such as about January 2000 and mid-2002.

 % =======================================================================

 \section{INPOP08 Fit}

 For the fit of INPOP08, MEX and VEX observations provided by ESA and described previously were added to the INPOP06 data sets used for its adjustment. See (Fienga et al. 2008) for a more detailed description of this data set. Observations deduced from the tracking of the Cassini spacecraft processed and provided by JPL (Folkner et al. 2008) were also added to the set of observations.

 Global adjustments of planet initial conditions, Earth-Moon mass ratios, Astronomical Unit, Sun oblateness J2 as well as 34 asteroid masses were fitted to obtain INPOP08. Obtained values are presented in table \ref{paramphy} and table \ref{newmass} as well as comparable values found in the literature. The fit procedures are discussed  in the following sections.

 \begin{table}
 \caption{Residuals obtained from INPOP06 and INPOP08}
 \begin{tabular}{l l l | l | l }
 \\
 \hline
 Planet & Data  & Nbr & INPOP06 & INPOP08 \\
 & [unit] & & 1$\sigma$& 1$\sigma$\\
 \hline
 {\bf{Mercury}} & & & & \\
 {\tiny{Direct radar}} & range [m] & 462 & 894 & 842 \\
 {\bf{Venus}} & & & & \\
 {\tiny{Magellan}}    & VLBI [mas] &    18&     2 &   2 \\
  {\tiny{Direct radar}}   & range [m] &   488&   1386 &  1376 \\
 {\tiny{VEX}} & range [m] & 15131 &   185.829 &  4.6 \\
 {\bf{Mars}} & & & & \\
 {\tiny{MGS/MO}} & range [m] & 10410&    5.954 &     1.57 \\
 {\tiny{MEX}} & range [m] &  6006&  13.56 &    2.07 \\
 {\tiny{Path}} & range [m] &    90&    7.63 &    12.46 \\
 {\tiny{Vkg}} & range [m] &  1245&    17.396 & 18.53 \\
 {\tiny{Mixed}} & VLBI [mas] &    96 &    0.4 &    0.4 \\
 {\bf{Jupiter}} & & & & \\
 {\tiny{Galileo}} & VLBI [mas]& 24 & 12 &      11 \\
 {\tiny{Optical}} & ra [arcsec]& 5616 & 0.343 &      0.343 \\
 {\tiny{Optical}} & de [arcsec]& 5534 & 0.332 &      0.338 \\
 {\bf{Saturn}} & & & & \\
 {\tiny{Optical}} & ra [arcsec]& 5598 & 0.347 &      0.346 \\
 {\tiny{Optical}} & de [arcsec]& 5573 & 0.312 &      0.311 \\
 {\tiny{Cassini}} & ra [mas] &   31&   5 &  4 \\
 {\tiny{Cassini}} & de [mas] &   31&   7 &  7 \\
 {\tiny{Cassini}} & range [m] & 31&   27324 & 22 \\
 {\bf{Uranus}} & & & & \\
 {\tiny{Optical}}  & ra [arcsec]& 3849 &    0.358   &      0.351 \\
 {\tiny{Optical}}  & de [arcsec]& 3835 &    0.366    &      0.361 \\
 {\bf{Neptune}} & & & & \\
 {\tiny{Optical}} & ra [arcsec]& 3898 &   0.368     &      0.361 \\
 {\tiny{Optical}} & de [arcsec]& 3879 &    0.360    &      0.358 \\
 {\bf{Pluto}} & & & & \\
 {\tiny{Optical}} & ra [arcsec]& 1023 &    0.170  &      0.170 \\
 {\tiny{Optical}}  & de [arcsec]& 1023 &    0.171  &      0.171 \\
 \hline
 \end{tabular}
 \label{omc}
 \end{table}

 \subsection{Global results}
 \subsubsection{Contribution of VEX data}
 \label{RGvex}
 The impact of VEX observations on the INPOP08 adjustment is very important.
 As one can see in Figure \ref{vexplot} and table \ref{omc}, INPOP08 provides a much more accurate Venus orbit thanks to the VEX input data. This improvement is mainly induced by the fit of planet initial conditions and also by the improvement of the modelisation of the asteroid perturbations on the inner planets.  A mean dispersion of 4 meters (1-sigma formal dispersion) with a bias of 1.5 meters is obtained after the fit,  which corresponds to the level of accuracy of the VEX observations. This is an improvement in the estimation of the Earth-Venus distance of about a factor 42 compared to the previous ephemeris, INPOP06, which was not fitted to VEX data. The large effect that one can see on figure \ref{vexplot} is actually induced by the propagation of the previous planetary ephemerides uncertainties on Earth and Venus orbits. Such ephemerides as INPOP06, but also DE405, were only fitted to direct radar observations on the Venus surface (see  Standish 1998b; Fienga et al. 2008) and to the VLBI data deduced from the tracking of the Venus mission Magellan in 1994. The Venus data sets were suffering from a big lack of observations since 1994 and from low accuracy of the direct radar  observations. The combination of both explains the spectacular improvement observed on Figure \ref{vexplot}.
 One can also notice in table \ref{omc} that even if the new Venus INPOP08 orbit induces significant changes in the values of VEX residuals compared to INPOP06 ones, the residuals obtained by comparison between this new orbit and the observations used previously for the INPOP06 fit are quite similar to those obtained with INPOP06.
   This means that the VEX data are consistent with the old ones and their addition in the fit does not degrade the ephemerides on a large time interval.\\

 \begin{figure*}
 \begin{center}
 \includegraphics[width=14cm]{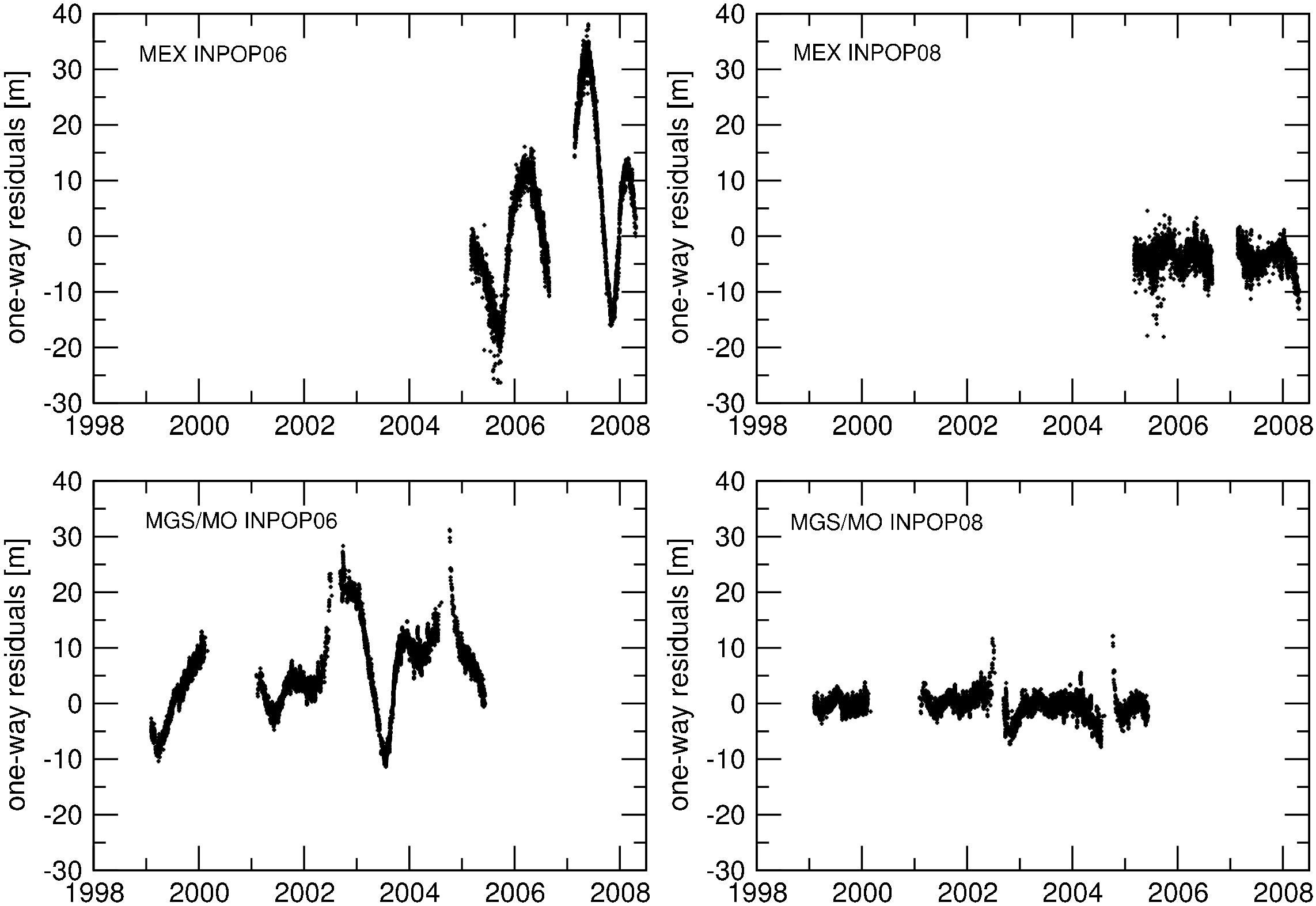}
 \caption{MEX and MGS/MO 1-way residuals with the new INPOP08 fitted to VEX, MEX and MGS/MO data (right-hand side plots) and INPOP06 fitted to all MGS/MO and a part of MEX data (left-hand side plots). The improvement of INPOP08 is mainly induced by changes in the asteroid mass determinations and the addition of MEX data. see section \ref{ringtxt} in the text. The y-axis unit is meter.}
 \label{mexplot}
 \end{center}
 \end{figure*}

 \subsubsection{Mars observations and asteroids}

 Using MEX tracking data and the INPOP06 Mars data set, we fitted the new asteroid ring modeling  as well as the selection of asteroids described in section \ref{ringtxt}. Compared to INPOP06, the ring model was modified to a non-static ring, 3 more asteroids were included in the list of main perturbers and more asteroids have their masses fitted individually in using a priori sigmas (Moyer 1971) of about 30$\%$ constraints to their initial values. The choice of the fitted asteroid masses was done in a way to have only positive masses even without the constraints.
In INPOP08, 34 asteroids have their masses fitted individually against 5 in INPOP06.

 As one can see in table \ref{omc} and figure \ref{mexplot}, this new approach and the input of information brought by the MEX data improve the ephemerides by reducing significantly the residuals. This improvement can be noticed with the new MEX data set which was not included in the previous solution, but also with the MGS/MO data set included both in INPOP08 and INPOP06 adjustments. For the MGS/MO residuals we have an improvement of the results of a factor 4. 
%Stability of INPOP08 can also be demonstrated over 30 years by noting that the new ephemerides give Viking residuals quite comparable to those obtained with INPOP06.

 Furthermore, as one can see on Figure \ref{mexplot} during the overlap period of the mission (from 2005.5 to 2006), the connection between MGS/MO and MEX data is done properly by INPOP08 with a bias of about 3.5 meters between MGS/MO and MEX. This  bias is in the error estimations presented in section \ref{mexvexdata} and the calibration done by ESA before the MEX launch. Even if a great improvement can be noticed on Figure \ref{mexplot}, signals still remain in INPOP08 residuals due to a still possible lack in asteroid mass determinations but also to systematics induced by solar conjunction gaps in the data sets.

\begin{figure}
 \begin{center}
 \includegraphics[width=9cm]{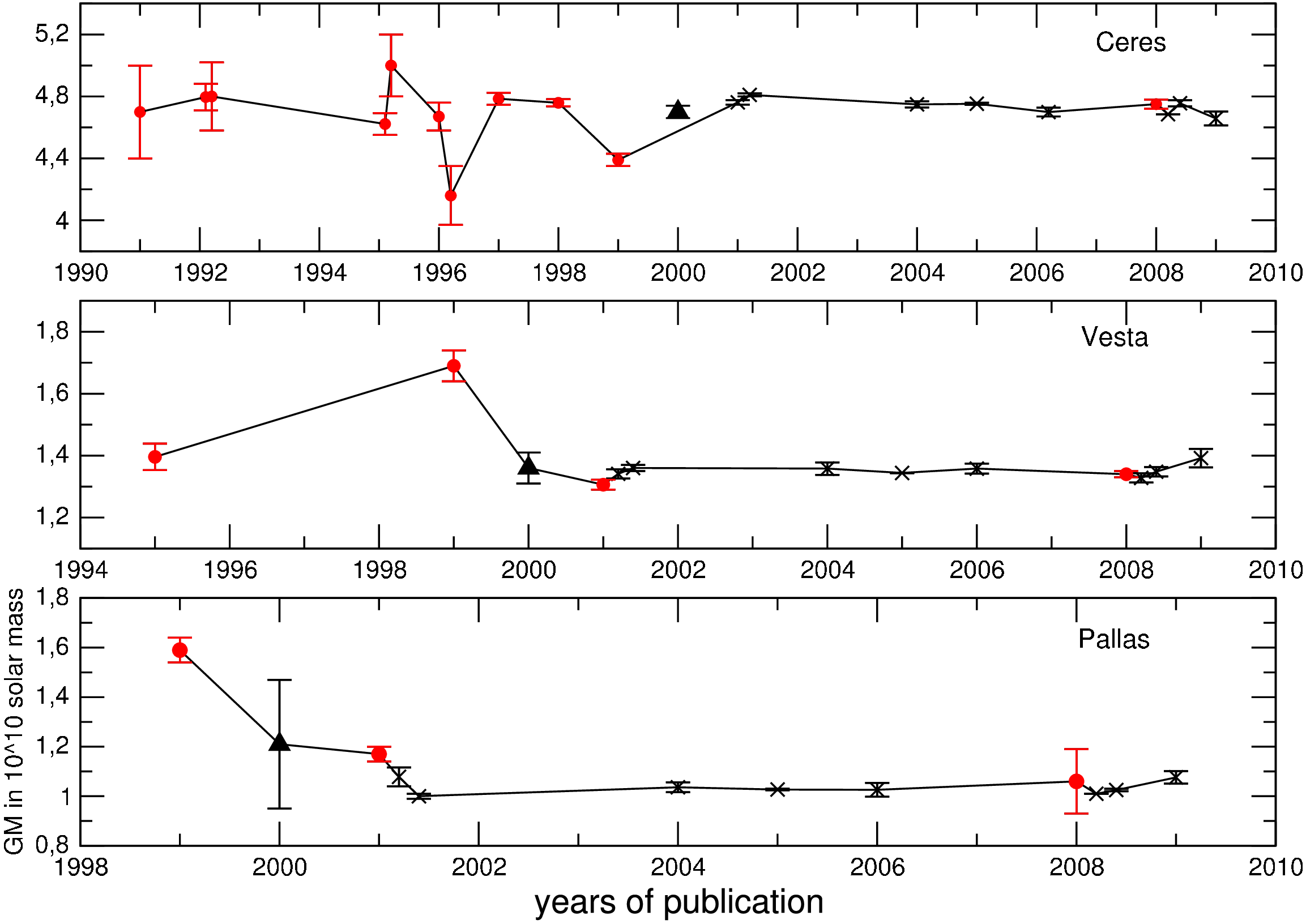}
 \caption{Estimations of Ceres, Pallas and Vesta masses found in the literature. The GMs (in unit of $10^{-10}$ solar mass y-axis) are plotted by years of publications (x-axis). The dark red circles represent the GMs estimated in using close-encounters as described by Baer and Chesley (2008). The crosses represent GMs fitted with the planetary ephemerides. The upper triangles represent masses deduced from radar observations as described by Michalak (2000). The last crosses corresponding to 2009 give values obtained by INPOP08.}
 \label{GMplot}
 \end{center}
 \end{figure}

 In Table \ref{paramphy} and \ref{newmass}, asteroid masses fitted in INPOP08 are compared to other published masses as well as values for taxonomic densities and physical characteristics of the asteroid ring. In the case of INPOP08, the ring mass and distance were not fitted during the global adjustment but during the construction of the ring (see section \ref{ringtxt}). Due to the direct correlation between the mass of the ring and its distance to the Sun, we decided to fix arbitrarily  the distance at 3.14 AU. If this distance is chosen at 2.8 AU, the mass of the ring is estimated to be $M_{ring} = 0.3\pm0.1 \times 10^{-10} M_{\bigodot}$ which is comparable with the GM estimated by DE414 and very close to the one given by INPOP06.
 In table \ref{paramphy}, the taxonomic densities estimated by INPOP08 stay of the same order as those estimated by DE414 or DE421 and the values of masses estimated for the 5 biggest asteroids are also in good agreement with those published previously.
 Figure \ref{GMplot} depicts the same conclusion.
 Indeed, on figure \ref{GMplot}, the values estimated since 1991 (Baer and Chesley 2008) of GMs of Ceres, Pallas and Vesta are plotted as well as the values provided by INPOP06, DE414, DE421 and INPOP08 which correspond to the four last points dated in 2008 and 2009. On this graph, are represented (using darked circles) asteroid masses estimated by close-encounters (see for instance Baer and Chesley 2008, Hilton 1999), asteroid masses obtained after radar imaging (Michalak 2000) and masses fitted during the global adjustments of planetary ephemerides marked as black crosses. The close-encounter method was used a lot in the 90s but due to the difficulty of the method (accurate observations of the perturbed objects to obtain very accurate estimation of its orbit and to survey the orbit during the close-encounters), the effort was not maintained during the last decade. On the other hand, the evolution of planetary ephemerides gives very stable estimations for the largest asteroids with uncertainties smaller than the close-encounters determinations. In 2008, (Baer and Chesley) published values established on the basis of several encounters and obtained values of masses for Ceres, Pallas and Vesta very close to the ones obtained by the planetary ephemerides. For Ceres and Vesta, the uncertainties of the close-ecounters determinations are at the same level of accuracy as planetary ephemerides ones. INPOP08 gives original values offset from the previous estimations but at the limit of the error bars.

 In table \ref{newmass}, are given the masses of the 28 asteroid, other than the 5 bigs given in table \ref{paramphy}, fitted individually in INPOP08. The estimated mass values are put in the table as their mean impact on the Earth-Mars distances (estimated over 20 years) decreases. For some of them, the INPOP08 masses agree in the limit of 30 \% with values estimated previously either by close-encounters methods (as for (16) Psyche, (52) Europa, (88) Thisbe, (8) Flora, (15) Eunomia, (18) Melpomene (19) Fortuna, and (21) Lutecia) or by adjustment of planetary ephemerides (as for (41) Daphne, (29) Amphitrite, (409) Aspasia, (704) Interamnia, or (532) Herculina). However, for some others, the differences from previous estimations can reach 500 \% as for the case of (9) Metis or even worse (as for (6) Hebe). We estimate to 50\% of estimations close to previously published masses (differences below 50 \%), 21\% of first estimations and 4 very unsufficent estimations (differences greater than 500 \%).

Such differences can be explained by the differences in the modelisation used to fit the data. For instance, by the differences in the data reduction, the weight used in the fit as well as by fitting individually a different number of asteroids and by absorbing the effect of all the other minor bodies by the fit of the taxonomic classes and the addition of a ring, one absorbs in the fit of individual masses dynamical effects which are not really caused by one particular asteroid but by a group of asteroids inducing effects very similar in amplitudes and periods.  The fitted mass for this particular asteroid represents then not only its gravitational potential but also the one induced by other asteroids. This remark can be illustrated by the unrealistic values of some densities deduced from masses and presented in table \ref{newmass} as well as the almost zero mass of the asteroid (747) Winchester, the perturbations induced by (747) being quite negligible or not clearly separated from another asteroid perturbations like in the case of (6) Hebe. 
In order to better estimate the real mass value of one particular asteroid, one could isolate the arc of Mars orbit where and when this asteroid has the biggest impact and try to estimate its mass only on that arc. Such an investigation is presented in (Somenzi et al. 2009).

 \subsubsection{Cassini normal points}

 Cassini normal points for Saturn were provided by JPL and are described in details by (Folkner et al. 2008).
 The impact of these data is not negligeable. They give very important informations about the Saturn orbit itself by providing very accurate estimations of the Earth-Saturn distances and geocentric angular positions of Saturn. Thanks to these, our knowledge of this orbit especially in term of distances from the Earth reduces from about 27 km to 22 meters at the epoch of Cassini normal points. 
%One needs to get such accurate points over a 30 years period of the Saturn orbit to improve it in its whole.}}
 %Secondly, in term of reference frames, the differences in angles provided by JPL improve the definition of the INPOP planetary reference frame in ICRF.  
%As one can see in table \ref{omc}, the impact is very important, especially for the Saturn orbit.

 \begin{table*}
 \caption{Physical parameters fitted in INPOP08. Other values deduced from planetary ephemerides
 are presented for comparison. EPM08 stands for (Pitjeva 2008). The uncertainties are given at 5-sigma for the GMs and 1 formal sigma for others. n/a stands for non-avalaible and NE as non-estimated. For AU, the presented values are the differences in meters between the fitted values and the AU value of the IERS conventions 2003, $\mathrm{AU_{IERS03}}$ = 149597870.691 km.
 }
 \begin{tabular}{l l l r r r l}
 \\
 \hline
 \\
 &Unit&
 %DE405 &
 %EPM2004 &
 DE414&
 DE421&
EPM 2008 & 
 INPOP06 & INPOP08
 \\
 \hline
 \\
 Mass of Ceres &
 $10^{-10} M_{\bigodot}$&
 %4.64&
 4.699 $\pm$ 0.028&
 4.685 &
4.712 $\pm$ 0.006 &
 4.756 $\pm$ 0.020&
 4.658 $\pm$ 0.045\\
 Mass of Vesta&
 $10^{-10} M_{\bigodot}$&
 %1.34&
 1.358 $\pm$ 0.016 &
 1.328  &
 1.344 $\pm$ 0.003&
 1.348 $\pm$ 0.015 &
 1.392 $\pm$ 0.015 \\
 Mass of Pallas&
 $10^{-10} M_{\bigodot}$&
 %1.05&
 1.026 $\pm$ 0.028&
 1.010  &
1.027 $\pm$ 0.007 &
 1.025 $\pm$ 0.005 &
 1.076 $\pm$  0.010\\
 Mass of Juno&
 $10^{-10} M_{\bigodot}$&
 %1.05&
 0.149 $\pm$ 0.015&
 0.116  &
n/a &
 NE &
 0.075 $\pm$ 0.015 \\
 Mass of Iris&
 $10^{-10} M_{\bigodot}$&
 %&
 0.060 $\pm$ 0.010 &
 0.060 &
n/a &
 0.058 $\pm$ 0.005 &
 0.050 $\pm$  0.010\\
 Mass of Bamberga&
 $10^{-10} M_{\bigodot}$&
 %&
 0.047 $\pm$ 0.007&
 0.048 &
n/a &
 0.046 $\pm$ 0.003 &
 0.056 $\pm$   0.004\\
 Mass of Ring&
 $10^{-10} M_{\bigodot}$&
 %&
 0.30 & NE &n/a &
 0.34 $\pm$ 0.1& 1.0 $\pm$ 0.3\\
 Distance of Ring&
 UA&
 %&
 2.8 &  NE & 
n/a &
 2.8  & 3.14 \\
 Density of the C class&
 &
 %1.8
 1.62 $\pm$ 0.07&
 1.09  &
n/a &
 1.56 $\pm$ 0.02 &
 1.54 $\pm$ 0.07 \\
 Density of the S class&
 &
 %2.4&
 2.08 $\pm$ 0.19&
 3.45&
n/a &
 2.18 $\pm$ 0.04 &
 1.94 $\pm$ 0.14\\
 Density of the M class&
 &
 %5.0&
 4.32 $\pm$ 0.37 &
 4.22  &
n/a &
 4.26 $\pm$ 0.12 &
 4.98 $\pm$ 0.50\\
 Sun J2&
 $10^{-7}$&
 %2 &
 2.34 $\pm$ 0.49 &
 2.0 &
n/a &
 2.46 $\pm$ 0.40 &
 1.82 $\pm$ 0.47  \\
 EMRAT &
 & 81.300568 & 81.300569 & 81.3005690 $\pm$ 0.0000001 & NE & 81.300540 $\pm$ 0.00005\\
 AU-$\mathrm{AU_{IERS03}}$ & m & 9.8 $\pm$ 0.15 & 8.6 $\pm$ 0.15 & 4.4 $\pm$ 0.10 & NE &  8.22 $\pm$ 0.11 \\
  \\
 \hline
 \end{tabular}
 \label{paramphy}
 \end{table*}

 \begin{table*}
 \caption{Asteroid GMs fitted in INPOP08. Other values deduced from planetary ephemerides
 are presented for comparison. Reference [1] stands for (Baer and Chelsey, 2008). [2] are values estimated in DE414 but not published with errorbars. They are directly extracted from the DE414 header. [3] stands for (Baer et al. 2008) and [4] for DE421 values as published by (Folkner et al. 2008). The given uncertainties are  formal accuracy given at 1-sigma. Column 4 and 5 are dedicated to physical characteristics of the objects: in column 4 are given radius in kilometers, and spectral type in column 5. In column 6, are given the densities obtained in using the given radius and the estimated masses given in column 2. Density errorbars are estimated with the given mass 1-sigma uncertainties and accuracies on radius when provided by the PDS website. The last column gives the maximum impact of each asteroid on the Earth-Mars distances over the 1998 to 2008 intervalle of time.
}
 \begin{tabular}{l l r l l l l}
 \\
 \hline
 \\
 Asteroid & INPOP08 &  Others  & r & type & $\rho$ & \\
& $10^{-11} M_{\bigodot}$ &  $10^{-11} M_{\bigodot}$ & km & & g.$cm^{-3}$ & [m]\\
 \hline
 \\
%3 Juno & 0.753 $\pm$ 0.02 & 1.49 $\pm$ 0.15 [1] & 117 $\pm$ 6 & S & 2.29 $\pm$ 0.3& 184\\
%&& 1.49 $\pm$ 0.15 [1]& & &&\\
%&& 1.54  [4]& & &&\\
 16 Psyche & 1.596 $\pm$ 0.032 & 1.29 $\pm$ 0.17 [1] & 126.6 $\pm$ 2.0 & M & 3.7 $\pm$ 0.1 & 138\\
&& 1.688 [4]& & &&\\
 29 Amphitrite &0.491 $\pm$ 0.09 & 1.00 $\pm$ 0.35 [1] & 106.1 $\pm$ 3.4 & S & 1.9 $\pm$ 0.4 & 118\\
&& 0.684 [4]& & &&\\
 14 Irene & 0.071 $\pm$ 0.01 & 0.2623 [2] & 76.0 $\pm$ 8.0 & S & 0.8 $\pm$ 0.3 & 69\\
& & 0.413 $\pm$ 0.073 [3] & & &&\\
&& 0.263 [4]& & &&\\
704 Interamnia & 1.623 $\pm$ 0.01 &  3.58 $\pm$ 0.42 [1] & 158.3 $\pm$ 2.0  & C & 1.9 $\pm$ 0.1& 65\\
&& 1.862 [4]& & & &\\
532 Herculina & 0.546 $\pm$ 0.01 & 0.667 [2] & 111.2 $\pm$ 2.0 & S & 1.9 $\pm$ 0.1& 63\\
&& 0.669[4]& & &&\\
 6 Hebe & 0.016 $\pm$ 0.011 & 0.759 $\pm$ 0.142 [1] & 92.5 $\pm$ 1.5 & S & 0.1 $\pm$ 0.1 & 63\\
 & & 0.255 [2] & & &&\\
&& 0.457 [4]& & &&\\
 52 Europa & 1.724 $\pm$ 0.100 & 0.976 $\pm$ 0.22 [1] & 151.25 $\pm$ 2.7 & C & 2.4 $\pm$ 0.2& 52\\
&& 1.023 [4]& & &&\\
9 Metis & 0.115 $\pm$ 0.100 & 1.03 $\pm$ 0.24 [1] & 95.0 $\pm$ 9.5 & S & 0.6 $\pm$ 0.5 & 49\\
 & & 0.600 [2] & & &&\\
&& 0.428 [4]& & &&\\
 19 Fortuna  & 0.202 $\pm$ 0.02 & 0.541 $\pm$ 0.008 [1] & 101.7  & C& 0.9 $\pm$ 0.1& 48\\
&& 0.350 [4]& & &&\\
128 Nemesis &0.169 $\pm$ 0.033 & &94.1 $\pm$ 2.0 & C & 1.0 $\pm$ 0.2 & 42\\
18 Melpomene & 0.091 $\pm$ 0.02 & 0.151 $\pm$ 0.051 [3] & 70.3 $\pm$ 1.5 & S & 1.2 $\pm$ 0.3& 40\\
&& 0.201 [4]& & &&\\
 15 Eunomia & 2.230 $\pm$ 0.018 & 1.68 $\pm$ 0.08 [1] & 127.7 $\pm$ 7.0 & S & 5.0 $\pm$ 0.7& 37\\
&& 1.239 [4]& & &&\\
88 Thisbe &0.863 $\pm$ 0.026 & 0.57 $\pm$ 0.18 [1] & 116.0 $\pm$ & C & 2.6 $\pm$ 0.1& 36\\
409 Aspasia  & 0.105 $\pm$ 0.003 &0.163 [4] &80.8 $\pm$ 4.0& C & 0.9 $\pm$ 0.1& 29\\
 216 Kleopatra & 0.353  $\pm$ 0.012 & 0.255 [2]&67.5 $\pm$  & M & 5.3 $\pm$ 0.2 & 27\\
&& 0.226 [4]& & &&\\
21 Lutetia & 0.1034 $\pm$ 0.03 & 0.129 $\pm$ 0.012 [3]& 47.9 $\pm$ 4.1 & M& 4.4 $\pm$ 1.3&24\\
&& 0.105 [4]& & &&\\
 31 Euphrosyne & 2.99 $\pm$ 0.68 & 1.54 [2] & 127.95 $\pm$ 5.5 & C & 6.7 $\pm$ 1.6& 23\\
& & 0.313 $\pm$ 0.059 [3] & & &&\\
&& 0.860 [4]& & &&\\
23 Thalia & 0.003 $\pm$ 0.001& &53.8 $\pm$ 2.2 & S & 1.06 $\pm$ 0.08& 23\\
&& 0.097 [4]& & &&\\
354 Eleonora & 0.488 $\pm$ 0.035 & 0.246  [2] &77.6 $\pm$ 4.2 & S & 4.9 $\pm$ 0.4& 22\\
&& 0.247 [4]& & &&\\
 192 Nausikaa & 0.137 $\pm$ 0.041 & 0.081 [4]& 51.6 $\pm$ 1.0 & S & 4.6 $\pm$ 1.4& 22\\
747 Winchester & 0.0004 $\pm$ 0.0002 & 0.148[4] &85.9 $\pm$ 1.5 & C & 3e-3 $\pm$ 1e-3 & 22\\
129 Antigone& 0.717 $\pm$ 0.014& &104.58 $\pm$ & M & 3.0 $\pm$ 0.1& 21\\
41 Daphne & 0.527 $\pm$ 0.05& 0.467 [2] & 87.00 $\pm$ 6.0 & C & 3.8 $\pm$ 0.7& 20\\
&&0.398[4]& & &&\\
173 Ino& 0.366 $\pm$  0.030 &&77.0 $\pm$ 1.6 & C & 3.8 $\pm$ 0.3& 20\\
89 Julia &0.359 $\pm$ 0.014& & 75.8 $\pm$ 1.5 & C & 4.0 $\pm$ 0.4& 15\\
 %3 Juno &  &  [2]   \\
8 Flora & 0.535 $\pm$ 0.005 & 0.178 [2] & 67.9 $\pm$ 1.0 & S & 8.1 $\pm$ 0.5 & 14\\
& & 0.426 $\pm$ 0.045 [3]& & &&\\
&& 0.178 [4]& & &&\\
  12 Victoria & 0.117 $\pm$ 0.003& &56.4 $\pm$ 1.5 & S & 3.1 $\pm$ 0.2& 13\\
 139 Juewa& 0.359 $\pm$  0.010 &0.142[4]&78.3 $\pm$ 1.5 & C  & 3.6 $\pm$ 0.10& 10\\
 \hline
 \end{tabular}
 \label{newmass}
 \end{table*}

 %\begin{figure}
 %\includegraphics[width=\figw]{ceres}
 %\caption{}
 %\label{GMplot}
 %\end{figure}

 \subsection{AU fixed and rescaled GM of the sun version}
\label{au}

 As a first step to a new generation of planetary ephemerides fitting the GM of the sun and using a fixed value of AU, an alternative version of INPOP08 was built. This version is based of the INPOP08 ephemeris presented in the previous section. However, instead of providing a value of AU fitted to observations, the AU is fixed to the (IERS 2003) value and the change is absorbed by the multiplication of all the initial conditions of planets and asteroids as well as all the masses including the GM of the sun by a factor equivalent to $\mathrm{AU_{fitted}}$ over $\mathrm{AU_{IERS03}}$. A new value for the GM of the Sun is then deduced and presented on Table \ref{AUGM}.
This version of INPOP08 is equivalent to the previous one but allows the next versions of planetary ephemerides to fit directly the GM of the sun instead of the AU. 
%Such conversion of the GM of the Sun adjustment instead of the AU will allow us to monitor possible variations in time of the $GM_\odot$.

\begin{table}
\caption{AU and GM of the sun values used for the construction of INPOP08 and INPOP08b. In the case of INPOP08, the AU is fitted and the mass of the Sun fixed to the DE405 one. In the case of INPOP08b, the AU is fixed to be equal to $\mathrm{AU_{IERS03}}$, and a new estimation of the mass of the Sun is deduced as described in section \ref{au}.} 
 \begin{tabular}{l l l  }
\hline
& AU &  GM$_\odot$  \\
& km & km$^{3}$.s$^{-2}$ \\
\hline
\\
 & Fitted & DE405 \\
INPOP08 & 149597870.69922& 132712440039.87900\\
\\
& IERS03 & Deduced \\
INPOP08b & 149597870.69100& 132712440039.878750\\
\hline
\end{tabular}
\label{AUGM}
\end{table}

\begin{figure*}
\begin{center}
\includegraphics[width=8cm]{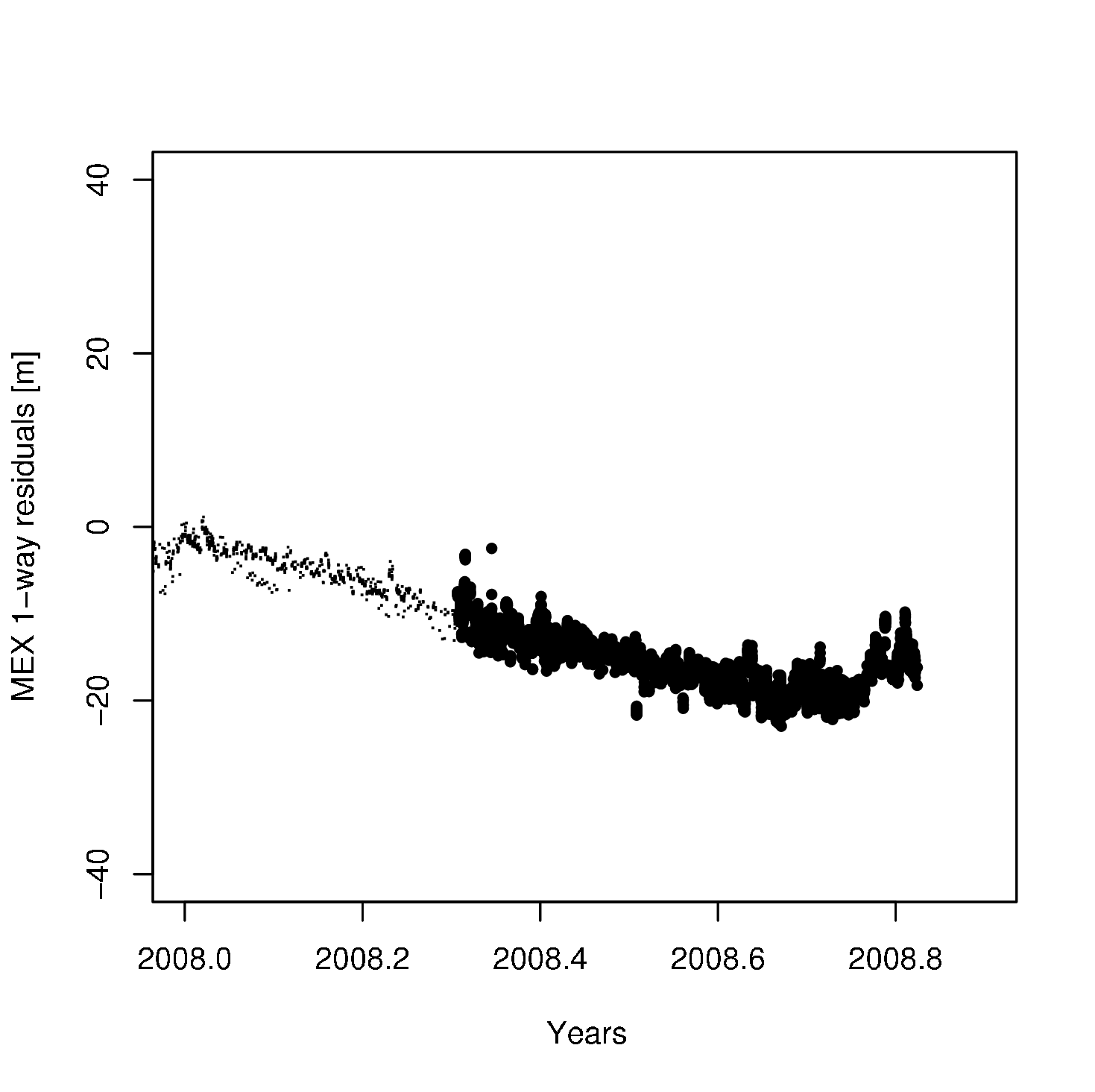}\includegraphics[width=8cm]{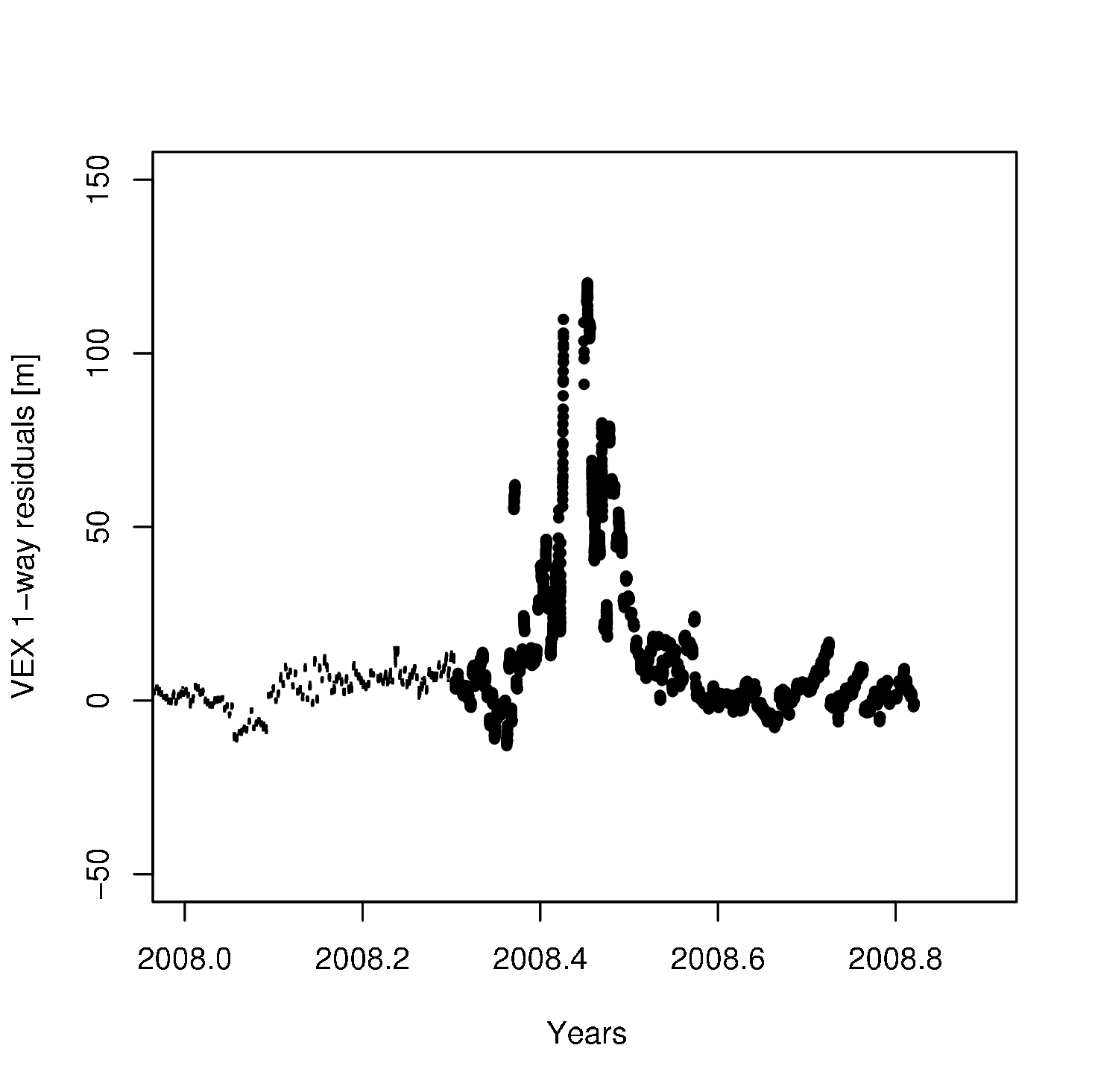}
\caption{Residuals obtained by comparisons between INPOP08 and MEX tracking data (left hand side) and VEX (right hand side) observations. The small dark points represent residuals of observations which were used in the fit of INPOP08 and the big dark points represent residuals of observations not used in the fit. The large slope of MEX residuals may possibily be improved in the future with better determination of asteroid masses.}
\label{omcout2}
\end{center}
\end{figure*}

 \section{Estimation of uncertainties}
\label{uncert}
 Often, users are asking for the accuracy of planet ephemerides. There are two methods to answer.
 The first one is to compare the ephemerides to observed positions not used in the fit, and preferably to observations situated out of the time interval of observations used for the adjustment. Such results give information to the users who directly deal with observations of planets or satellites: what accuracy in millarcseconds can we predict the position of such a planet in right ascension and declination, what accuracy in meters obtained for the estimation of Earth-Mars distances etc... . In section \ref{omcout}, we present such an extrapolation of INPOP08 compared to observations not used in the fit and usually outside the fit interval.

 However, for all the other users, the comparison to observations is interesting only if they correspond to their specific use. In general, users prefer to have a more theoretical estimation such as the uncertainties in the barycentric positions and velocities of the Earth. To answer this question, we use comparisons between different ephemerides    to estimate the level of variations induced by an improvement of modeling or adjustment (section \ref{compeph}).

 \subsection{Extrapolation tests}
 \label{omcout}

 %\subsubsection{Inner planets}

 Two sets of observations of MEX and VEX were kept out from the fit procedure as test sample of INPOP08.
 These sets were obtained from April 2008 to September 2008. Differences in meters obtained by comparisons
 between the observed and INPOP08 distances can be found in figure \ref{omcout2}.
 For Mars, the linear drift observed is in the order of error expected (about 20 meters over half a year) and is mainly due to asteroid modeling and adjustment.
 
 For Venus, the distribution is noisier, but in good continuity with the postfit residuals obtained after the INPOP08 fit and presented in Figure \ref{omcout2}. The biggest noticeable effect on these residuals is induced by the VEX solar conjunction that occured on June 9th 2008. This solar conjunction has introduced unmodeled noise into the doppler and ranging data of VEX during a period of few days (between the 06 to 11 June inclusive) during when the spacecraft tracking was stopped. The peak that can be noticed on Figure \ref{omcout2} is then mainly induced by this conjunction.

 \subsection{Uncertainties by comparisons with other ephemerides}
 \label{compeph}

\begin{figure*}
\begin{center}
\includegraphics[scale=0.5]{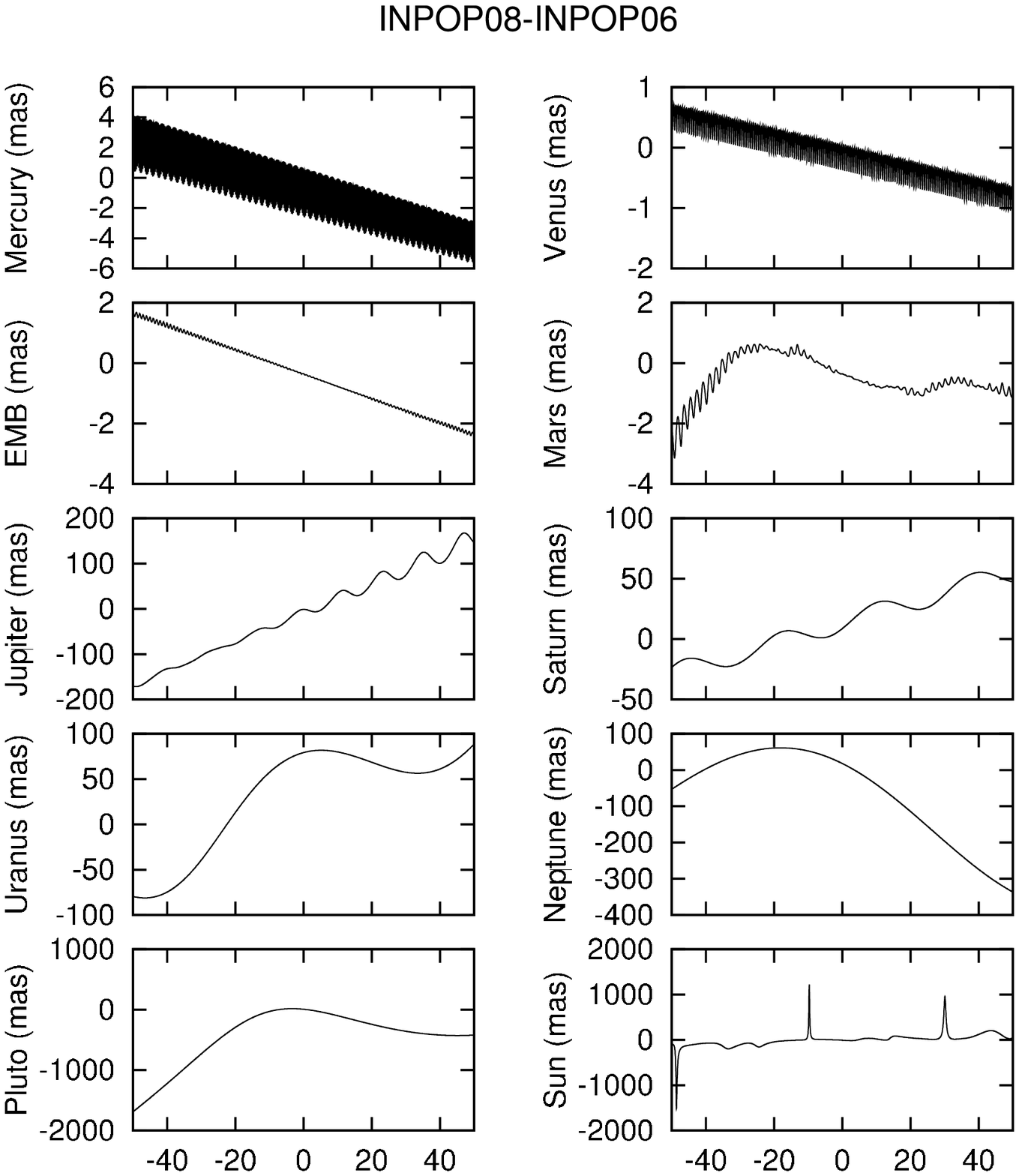}\includegraphics[scale=0.5]{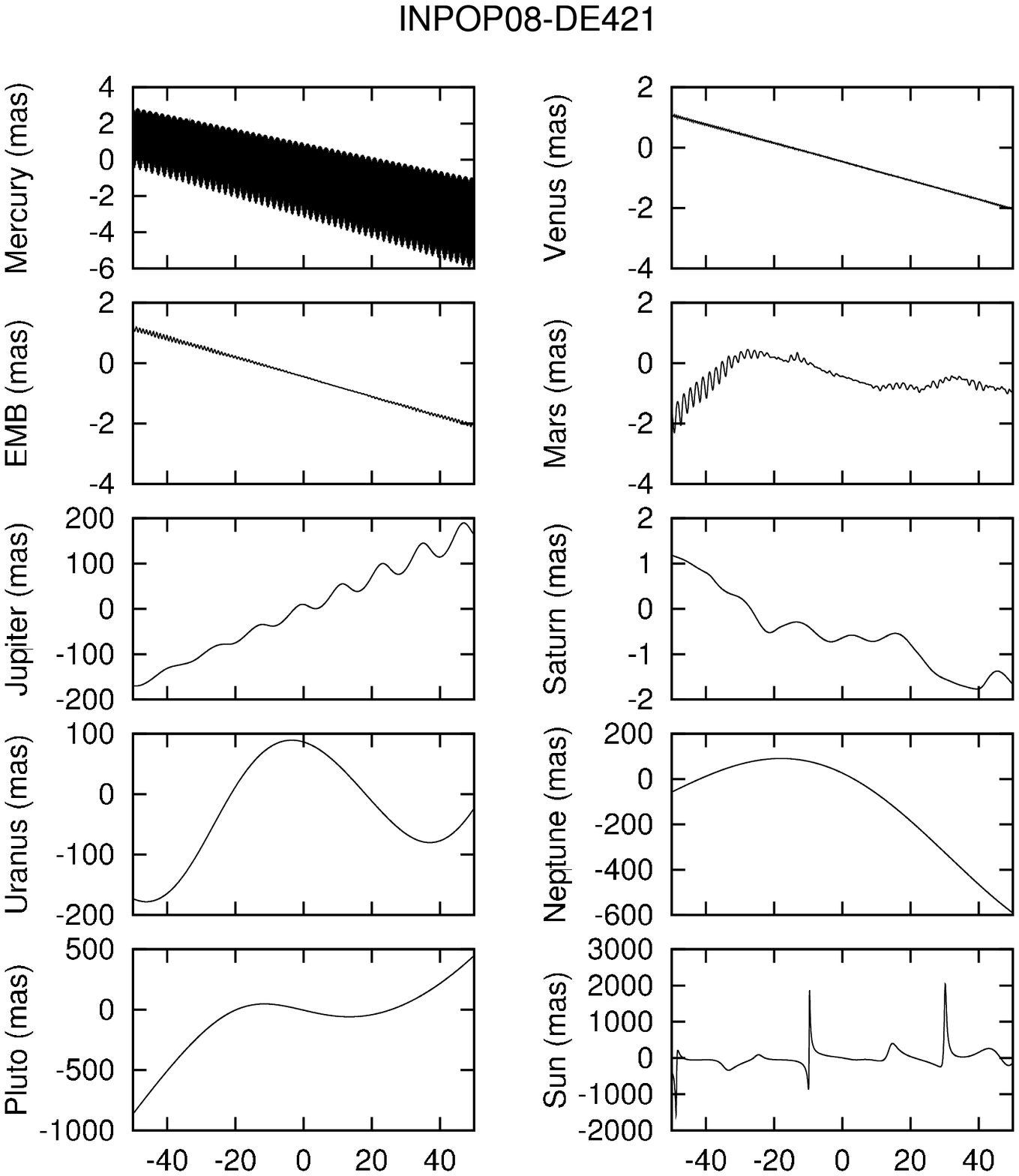}
\caption{Differences between INPOP08 and INPOP06 (left-hand side) and between INPOP08 and DE421 (right-hand side). For each planet, comparisons are made on the heliocentric longitude (barycentric for the Sun), expressed in the ecliptic frame. Units are milliarcseconds for Y-axis and years around J2000 for X-axis. }
\label{long}
\end{center}
\end{figure*}

 Another way to estimate the uncertainties of planet orbit is to compare several
 ephemerides. By such comparisons, one estimates first the impact of the differences in dynamical modeling of each ephemeris and in adjustment to observations: the ephemerides could have been fitted over different sets of observations with different sets of weight and with different sets of parameters.
Table \ref{inpop0806} presents the maximum differences in barycentric positions and velocities between INPOP08 and INPOP06. 
Table  \ref{inpop08de421}  presents the corresponding maximum differences between   INPOP08 and DE421. Two intervals of time are considered to estimate these differences: interval A from 1990 to 2010 and interval B from 1900 to 2050.
Furthermore, plots of heliocentric differences in ecliptic longitudes obtained by comparisons between the same ephemerides are presented in figure \ref{long}. 
The use of Cassini normal points in DE421 and INPOP08 make the differences between the two ephemerides decrease significantly for the longitude of Saturn, as well as for the other outer planets.

If one consider that these differences  give  a good estimation of their actual accuracies, then thanks to the Cassini data, we can see the improvement of the orbit at the epoch of these observations.

 Limited improvements of the outer planet orbits will be possible by the addition of more accurate data in the fit, mainly tracking data of spacecraft orbiting or cruising one of these systems, such as the  New Millenium mission.
Important improvements will be obtained after having significant portion of orbit covered by accurate observations which will require at least five years of continuous observations for Jupiter orbit.

For Mars and the EMB, one can see on figure \ref{long} the differences are also reduced between DE421 and INPOP08 compared to the differences obtained between INPOP06 and INPOP08.
As well for the EMB in table 5, if one uses the differences between ephemerides as a tool to estimate ephemerides accuracy, the uncertainties on its barycentric positions are about 900 meters over 10 years and 4 kilometers over 100 years, and about 10 meters per day to 50 meters per day for its barycentric velocities.

 \begin{table}
 \caption{Maximum differences between INPOP06 and INPOP08 estimated on interval A from 1990 to 2010 and on interval B from 1900 to 2050.}
 \begin{center}
 \begin{tabular}{r r r | r r }
  \hline
  & \multicolumn{2}{c}{Barycentric } & \multicolumn{2}{c}{Barycentric}\\
 & \multicolumn{2}{c}{Positions ($m$)} & \multicolumn{2}{c}{Velocities ($m/d$)}\\
% & 10 yrs & 100 yrs & 10 yrs & 100 yrs \\
Interval & A & B & A & B \\
  \hline
  Mercury          & 3500  & 4600 & 260 & 280    \\
  Venus            & 1700  & 2500 & 37 & 41    \\
  EMB              & 760  & 4000 & 11 & 48    \\
  Mars             & 980  & 14000 & 9 & 140    \\
  Jupiter          & 240000  &  1300000 & 350 & 1900    \\
  Saturn           & 320000  & 560000 & 161 & 320    \\
  Uranus           & 1800000  & 1800000 & 290 & 410    \\
  Neptune          & 2400000  & 7500000 & 320 & 620    \\
  Pluto            & 7500000  & 140000000 & 1700 & 7200    \\
  \hline
  \end{tabular}
 \end{center}
 \label{inpop0806}
 \end{table}

 \begin{table}
 \caption{Maximum differences between DE421 and INPOP08 estimated on interval A from 1990 to 2010 and on interval B from 1900 to 2050.}
 \begin{center}
 \begin{tabular}{r r r | r r }
  \hline
  & \multicolumn{2}{c}{Barycentric } & \multicolumn{2}{c}{Barycentric}\\
 & \multicolumn{2}{c}{Positions ($m$)} & \multicolumn{2}{c}{Velocities ($m/d$)}\\
 Interval & A & B & A & B \\
  \hline
  Mercury          & 3900      & 4400 & 290 &  310  \\
  Venus            & 740       & 3200 & 13 &  39  \\
  EMB              & 870       & 3900 & 11 &  36  \\
  Mars             & 1300      & 9200 & 10 &  82  \\
  Jupiter          & 210000    & 1300000 & 290 &  2000  \\
  Saturn           & 35000     & 40000 & 22 &  23  \\
  Uranus           & 1400000   & 2500000 & 190 &  390  \\
  Neptune          & 3700000   & 14000000 & 480 &   1200  \\
  Pluto            & 4300000   & 120000000 &  830 &  6700  \\
  \hline
  \end{tabular}
 \end{center}
 \label{inpop08de421}
 \end{table}

 \section{Perspectives and conclusions}
\label{future}
In this article, INPOP08 was fully described as a new 4-D planetary ephemeris.

New modeling of asteroid perturbations were introduced and as described by
 Kuchynka et al. (2009), improvements in the methods of asteroid mass determinations are possible.
Introduction of these new methods will be done in the next INPOP version, leading to an improvement of the extrapolation capabilities of the ephemerides as well as more realistic determination of asteroid masses.

New sets of observations were used for the construction of INPOP08: the Mars Express and Venus Express tracking data processed by ESA and normal points deduced from the Cassini mission.
Thanks to these data, large improvements were made in the determination of the orbits of Venus and Saturn. They open doors to new tests of gravity which were hitherto limited to  only Earth-Mars distance accurate observations.
These tests and their results will be presented in a dedicated paper very soon.

However, as  was already stressed in section \ref{RGvex}, VEX tracking data and Saturn Cassini normal points have a crucial role to play in such investigations.

The LLR data have been used for the adjustment of the lunar orbit and libration in INPOP08 (Manche et al. 2007). The fit to these data as well as the complete lunar model will be described in a forthcoming paper.

The INPOP08 planetary ephemeris is available for users on the INPOP website 
(www.imcce.fr/inpop) as Chebychev polynomials coefficients interpolated via provided routines in C and fortran. Positions and velocities of the nine planets, the Sun, the moon libration and the 3 Euler angles of the Earth rotation can be extracted as well as the  TT-TDB relation at any time.
An alternative version with rescaled GM of the sun and AU fixed is also provided on request.

In the next version, we plan to adjust directly the GM of the sun instead of the AU. Simulations of Mercury missions data and use of fly-by data will be introduce in order to evaluate the improvements in terms of accuracy limits in the determination of PPN parameters and J2.

\begin{acknowledgements}
We thank S. Klioner for multiple dicussions during this work and M. Standish for constructive comments. 
This work was supported by CNES under contract 05/CNES//00-DCT 094, by the CS of Paris Observatory, and by PNP-CNRS.
\end{acknowledgements}


\begin{thebibliography}{}

%\bib{}{Anderson, J.D., Jurgens, R.F., Lau, E.L., Slade, M.A., III, Schubert, G.}{1996}{Shape
%and Orientation of Mercury from Radar Ranging Data.}{Icarus}{124}{690}{697}

%\bib{}{Anderson, J.D., Laing, P.A., Lau, E.L., Liu, A.S., Nieto, M. M., Turyshev, S.G }{2005}{Study of the anomalous acceleration of Pioneer 10 and 11}{arXiv}{0104064}{}{}

\bib{2008CeMDA.100...27B}{Baer, J., Chesley, S.~R.}{2008}{Astrometric masses of 21 asteroids, and an integrated asteroid ephemeris}{Celestial Mechanics and Dynamical Astronomy}{100}{27}{42}

\bib{}{Baer, J., Milani, A., Chesley, S., Matson, R.~D.}{2008}{An Observational Error Model, and Application to Asteroid Mass Determination}{AAS/Division for Planetary Sciences Meeting Abstracts}{09-28}{40}{52.09}

%\bib{}{Bertotti, B., Iess, L., and Tortora, P.}{2003}{A test of general relativity using radio links with the Cassini spacecraft}{Nature}{425}{374}{376}

%\bib{} {Bou\' e, G.,  laskar, J.}{2006}{Precession of a planet with a satellite}
 %{Icarus}{185}{vol. 2}{312}

%\bib{}{Bretagnon, P., Francou, G., Rocher, P., Simon, J.-L.}{1998}{SMART97:
%a new solution for the rotation of the rigid Earth}
%{\aap}{329}{329}{338}

\bib{} {Budnik, F., Morley, T. A. and Mackenzie, R. A.}{2004}{ESOC’s System for Interplanetary
Orbit Determination}{ESA SP}{548}{387}{392.}

%\bib{} {Capitaine, N., Wallace, P.~T., Chapront, J.}{ 2005}{Improvement of the IAU 2000 precession
%model.}{\aap}{432}{355}{367}

%\bib{} {Chapront, J., Chapront-Touz{\'e}, M. and Francou, G.}{2002}{A new determination of lunar orbital parameters, precession constant and tidal acceleration from LLR measurements}{\aap}{387}{700}{709}

%\bib{} {Chicarro}{A}{2002}{Mars Express Mission and Astrobiology}{Solar System Research}{vol 36}{6}{487}{491}

%\bib{}{Cox, C.M., Chao, B.}{2002}{Detection of a Large-Scale Mass Redistribution in the Terrestrial System Since 1998} {Science} {297}{831}{833}

 \bibitem{} Damour, T., \& Vokrouhlick{\'y}, D.\ 1995, Physical Review D, 52, 4455

\bib{}{Damour, T.,Soffel, M.,  Xu, C.}{1991}{General-relativistic celestial mechanics. I. Method and definition of reference systems}{\prd}{43}{3273}{3307}

 \bib{} {Fairhead, L. Bretagnon, P.}{1990}{An analytical formula for the time transformation TB-TT}{\aap}{ 229}{240}{247}

 \bib{} {Fienga, A., Manche, H., Laskar, J.,Gastineau, M.}{2008}{INPOP06: a new numerical planetary ephemeris}{\aap}{477}{315}{327}

%\bib{}{Folkner, W.M.}{1992}{Preliminary results from VLBI measurement of Venus on September 12, 1990}{JPL IOM}{335.1}{92}{25.}

%\bib{}{ Folkner, W.M.}{1993}{Results from VLBI measurement of Venus on March 29, 1992}{ JPL IOM}{335.1}{93}{22}

%\bib{}{Folkner, W.M.}{1994a}{Results from VLBI measurements of Venus on December 22 and December 23, 1991}{ JPL IOM}{ 335.1}{94}{006}

%\bib{}{Folkner, W.M. }{1994b}{Results from VLBI measurement of Venus on April 1, 1994}{ JPL IOM}{335.1}{94}{014}

%\bib{1997JGR...102.4057F} {Folkner, W.M., Kahn,
%R.D., Preston, R.A., Yoder, C.F., Standish, E.M., Williams, J.G.,
%Edwards, C.D., Hellings, R.W., Eubanks, T.M., Bills, B.~G.}{1997}{Mars
%dynamics from Earth-based tracking of the Mars Pathfinder lander.}{\jgr}{102}{ 4057}{4064.}

\bib{fol8}{Folkner, W.M.}{2008}{JPL planetary and Lunar ephemerides, DE421}{JPL Interoffice Memorandum}{343R}{08}{003}{}

\bib{IF}{Irwin, A.~W., Fukushima, T.}{1999}{A numerical time ephemeris of the Earth}{\aap}{348}{642}{652} 

%\bib{} {Fricke, W.}{1971}{A Rediscussion of Newcomb's Determination of Precession.}{\aap}{13}{298}{308}

%\bibitem{} Hairer, E., Nordsett, S. , Wanner, G.: 1993, Solving Ordinary Differential Equations, Vols. I, Second Edition,  {\it Springer--Verlag}

\bib{}{Hilton, J.~L.}{1999}{US Naval Observatory Ephemerides of the Largest Asteroids}{\aj}{117}{1077}{1086} 

\bib{}{IERS Conventions 2003,}{2004}{McCarthy, D.D., Petit. G.}{IERS Technical Note}{32} {Frankfurt am Main: Verlag des Bundesamts fur Kartographie und Geodesie}{127}


\bibitem{KlionerPrivate}{Klioner, S.~A.}{2007}{private communication.}

 \bibitem{2008IAUS..248..356K} {Klioner, S.~A.\ 2008, IAU Symposium, 248, 356 }

%\bib{2002Icar..158...98K} {Krasinsky, G.A., Pitjeva, E.V., Vasilyev, M.V., Yagudina, E.I}{2002}{Hidden Mass in the
%Asteroid Belt.}{Icarus}{158}{98}{105}

\bib{konopliv}{Konopliv, A.~S., Yoder, C.~F., Standish, E.~M., Yuan, D.-N., Sjogren, W.~L.}{2006}{A global solution for the Mars static and seasonal gravity, Mars orientation, Phobos and Deimos masses, and Mars ephemeris}{Icarus}{182}{23}{50}


\bib{}{Kuchynka, P., Laskar, J., Fienga, A., Manche, H., Somenzi, L.}{2009}{Improving the asteroid perturbations modeling in planetary ephemerides}{JOURNEES-2008/ Astrometry, Geodynamics and Astronomical Reference Systems}{ed. M. Soffel, N. Capitaine. }{Dresden}{in press}

 %\bib{Lambeck}{Lambeck, K.}{1988}{Geophysical geodesy}{Clarendon Press}{}{}{}


%\bib{2005CeMDA..91..351L}{Laskar, J}{2005}{Note on the Generalized Hansen and Laplace Coefficients.}{Celest. Mech.}{91}{351}{356}

%\bib{}{ Le Poncin-Lafitte, C., Manche, H., Fienga, A., Laskar, J.}
%{2006}{Consistency between General Relativity and planetary ephemerides}
%{private communication}{}{}{}

%\bib{}{Lindegren, L. Perryman, M.A.C.}{1998}{Microarcsec astrometry: the GAIA mission.}{Joint Discussion 14 of the XXIIIrd General Assembly of the IAU: The first results of Hipparcos and Tycho} {vol 14}{581}{582}

\bib{}{Manche, H., Bouquillon S., Fienga A., Laskar J., Francou G.}{2007}{Towards INPOP07, adjustments to LLR data}{JOURNEES-2007/The Celestial Reference Frame for the Future }{ed. N. Capitaine. }{Paris}{}

%\bibitem{}Markstein, P.: 2000, IA-64 and Elementary Functions, {\it Prentice Hall, New Jersey}

%\bibitem{}{Mathews, P.M., Herring, T.A., Buffet, B.A. }{2002}{Modeling of nutation
%     and precession : New nutation series for nonrigid Earth and
%      insights into the Earth's interior}{ J.Geophys.Res.}{ 107}{ B4}{}

\bib{}{Michalak, G.}{2000}{Determination of asteroid masses I}{\aap}{360}{363}{374}

 \bib{}{Moyer, T.}{2000}{Formulation for Observed and Computed Values of Deep Space Network
 Data Types for Navigation}{ed. Joseph H. Yuen}{John Wiley \& Sons}{}{}

\bib{}{Moyer, T.}{1971}{Mathematical Formulation of the double precision orbit determination}{Technical report 32-1527 }{JPL}{May 15}{1971}

 \bib{}{Morley, T.}{2006a}{2006b}{2007a}{2007b}{MEX data release}{private comm.}

 \bib{}{Morley, T. and Budnik, F.}{2007}{Effects on Spacecraft Radiometric Data at Superior
Solar Conjunction}{Proceedings 20th International Symposium on Space Flight Dynam-
ics}{Annapolis}{MD}{USA.}

% \bib{}{Newhall, X. X., Williams, J. G.}{1997}
% {Estimation of the Lunar physical librations}{Celestial Mechanics and Dynamical Astronomy} {66}{21}{30}


% \bib{2005SoSyR..39..176P} {Pitjeva, E.V.}{2005}{ High-Precision Ephemerides of
% Planets{\mdash}EPM and Determination of Some Astronomical Constants}{ Solar System Research}{ 39}{176}{186}

% \bib{2006IAUJD..16E..55P} {Pitjeva, E.V.}{2006.}{
% Limitations on Some Physical Parameters from Position Observations of
% Planets.}{Nomenclature, Precession and New Models in Fundamental Astronomy}
% {26th meeting of the IAU}{ Joint Discussion 16}{ 22-23 August 2006, Prague,}

\bib{}{Pitjeva, E. V.}{2008}{Ephemerides EPM2008: the updated model, constants, data}{JOURNEES-
2008: Astrometry, Geodynamics and Astronomical Reference Systems}{ed. M. Soffel, N. Capitaine. }{Dresden}{in press}

%\bib{1994Icar..112...27R} {Rappaport, N., Plaut, J.J.}{1994}{A 360-degree and -order model of Venus topography.}{Icarus}{112}{27}{33}

 %\bib{}{Somenzi, L.}{2007}{Multiarc analysis fitting}{private comm.}{}{}{}

%\bib{}{Shapiro, S.S., Davis, J.L., Lebach, D.E., and Gregory, J.S.}{2004}{Measurement of the solar gravitational deflection of radio waves using geodetic very-long-baseline interferometry data, 1979–1999}{ Phys. Rev. Lett.}{92}{121101}{}

\bib{}{Soffel, M., Klioner, S. A., Petit, G., Wolf, P., Kopeikin, S. M., Bretagnon, P., 
Brumberg, V. A., Capitaine, N., Damour, T., Fukushima, T., Guinot, B., Huang, T.-Y.,
Lindegren, L., Ma, C., Nordtvedt, K., Ries, J. C., Seidelmann, P. K., Vokrouhlický, D.,
Will, C. M., Xu, C.}{2003}{The IAU 2000 Resolutions for Astrometry, 
Celestial Mechanics, and Metrology in the Relativistic Framework: Explanatory Supplement}
{Astron. J.}{126}{2687}{2706}

\bib{}{Somenzi, L., Fienga, A., Laskar, J.}{2009}{Determination of asteroid masses through the short-arc analysis}{\aap}{submitted}{}{}

\bib{}{Standish, E.M.}{1998a}{Time scales in the JPL and CfA ephemerides}{\aap}{336}{381}{383}

\bib{STAN98}{Standish, E.M.}{1998b}{JPL planetary and Lunar ephemerides, DE405/LE405}{JPL Interoffice Memorandum}{312.F}{98}{048}

\bib{STAN01}{Standish, E. M., Williams, J.G.}{2001}{Orbital Ephemerides on Sun, Moon and planets}{\url{ftp://ssd.jpl.nasa.gov/pub/eph/planets/ioms/}}{ExplSupplChap8.pdf}{}{}

\bib{STAN06}{Standish, E. M.}{2006}{JPL Planetary, DE414}{JPL Interoffice Memorandum}{343R-06-002}{1}{8}

 %\bib{} {Svedhem, H,Titov, D.~V,McCoy, D,Lebreton, J.-P. , Barabash, S,Bertaux, J.-L,Drossart, P,Formisano, V., H{\"a}usler, B,Korablev, O,Markiewicz, W.~J. , Nevejans, D,P{\"a}tzold, M, Piccioni, G., Zhang, T.~L,Taylor, F.~W,Lellouch, E,Koschny, D., Witasse, O,Eggel, H,Warhaut, M,Accomazzo, A. , Rodriguez-Canabal, J,Fabrega, J,Schirmann, T. , Clochet, A,Coradini, M.,}{2007}{Venus Express{\mdash}The first European mission to Venus}{\planss}{55}{1636}{1652}

\bib{}{Tedesco, E.~F., Cellino, A., \& Zappal{\'a}, V.}{2005}{The Statistical Asteroid Model. I. The Main-Belt Population for Diameters Greater than 1 Kilometer}{\aj}{129}{2869}{2886}

%\bib{}{Wahr, J.M.}{1981}{The forced nutations of an elliptical, rotating, elastic, and oceanless Earth}{Geophys. J. Roy. Astron. Soc.}{64}{705}{}

%\bib{}{Wallace, P.T.}{ 2004}{SOFA software support for IAU 2000}{ American Astronomical Society Meeting}{204} {28.02}{ May 2004.}

%\bib{}{Will, C. M.}{2006}{The Confrontation between General Relativity and Experiment}{Living Rev. Relativity}{ 9} {3}{}

%\bib{1994AJ....108..711W} {Williams, J.G.}{1994}{
%Contributions to the Earth's obliquity rate, precession, and nutation.}{
%\aj}{108}{711}{724}



%\bib{}{Yoder, C.F., Dickey, J.O., Schultz, B.E., Eanes, R.J., Tapley, B.D.}{1983}
%{Secular variation of Earth's gravitational harmonic $J_2$ coefficient from LAGEOS and non-tidal acceleration of Earth rotation}{Nature}{303}{391}{412}



 \end{thebibliography}
 \end{document}